\RequirePackage{rotating} 
\documentclass[fleqn,usenatbib]{mnras}

\usepackage{newtxtext,newtxmath}

\usepackage[T1]{fontenc}

\DeclareRobustCommand{\VAN}[3]{#2}
\let\VANthebibliography\thebibliography
\def\thebibliography{\DeclareRobustCommand{\VAN}[3]{##3}\VANthebibliography}

\usepackage{letltxmacro}
\LetLtxMacro\oldttfamily\ttfamily
\DeclareRobustCommand{\ttfamily}{\oldttfamily\csname ttsize\endcsname}
\newcommand{\setttsize}[1]{\def\ttsize{#1}}%

\usepackage{caption}
\usepackage{graphicx}	

\usepackage{newtxtext,newtxmath}
\usepackage{mathtools}
\usepackage{amsmath}
\usepackage{dblfloatfix}

\makeatletter
\newcommand\notsotiny{\@setfontsize\notsotiny\@vipt\@viipt}
\newcommand\betweenscriptfootnote{\@setfontsize\notsotiny{7}{8}}
\makeatother

\title[M30 (NGC 7099) in the \textit{FUV}]{Far-ultraviolet investigation into the galactic globular cluster M30 (NGC 7099): II. Potential X-ray counterparts and variable sources}

\author[S. Mansfield et al.]{
Santana Mansfield$^{1}$\thanks{Email: smansfield@astro.uni-bonn.de},
Andrea Dieball$^{1}$,
Pavel Kroupa$^{1,2}$,
Christian Knigge$^{3}$,
David R. Zurek$^{4}$,
\newauthor
\ Michael Shara$^{4}$,
and Knox S. Long$^{5}$
\\
$^{1}$Helmholtz-Institut f\"ur Strahlen- und Kernphysik (HISKP), Universit\"at Bonn, Nu\ss allee 14-16, D-53115 Bonn, Germany\\
$^{2}$Astronomical Institute, Faculty of Mathematics and Physics, Charles University in Prague, V  Hole\v{s}ovi\v{c}k\'ach 2, CZ-18000 Praha, Czech Republic\\
$^{3}$School of Physics and Astronomy, University of Southampton, Highfield, Southampton SO17 1BJ, UK\\
$^{4}$Department of Astrophysics, American Museum of Natural History, New York, NY 10024, USA\\
$^{5}$Space Telescope Science Institute, 3700 San Martin Drive, Baltimore, MD 21218, USA\\
}

\date{Accepted XXX. Received YYY; in original form ZZZ}

\pubyear{2022}

\begin{document}
\label{firstpage}
\pagerange{\pageref{firstpage}--\pageref{lastpage}}
\maketitle

\begin{abstract}
We present a far-ultraviolet (\textit{FUV}) study of the globular cluster M30 (NGC 7099). The images were obtained using the Advanced Camera for Surveys (ACS/SBC, F150LP, \textit{FUV}) and the Wide Field Planetary Camera 2 (WFPC2, F300W, \textit{UV}) on board the \textit{Hubble Space Telescope (HST)}. We compare the catalogue of $FUV$ objects to ten known X-ray sources and find six confident matches of two cataclysmic variables (CVs), one RS CVn, one red giant with strong $FUV$ emission and two sources only detected in the $FUV$. We also searched for variable sources in our dataset and found a total of seven blue stragglers (BSs), four horizontal branch (HB) stars, five red giant branch stars, 28 main sequence stars and four gap objects that demonstrated variability. One BS star is a known W-UMa contact binary, one of the gap objects is a known CV identified in this work to be a dwarf nova, and the three other gap sources are weak variables. The periods and positions of two of the variable HB stars match them to two previously known RR Lyrae variables of types RRab and RRc.

\end{abstract}



\setttsize{\small}

\section{Introduction}

The high density of stars in the cores of globular clusters allows frequent interactions which produces various exotic sources such as blue straggler (BS) stars which are located at brighter magnitudes in the colour magnitude diagram (CMD) than the main sequence (MS) turnoff. BS stars are typically thought to have higher masses than MS stars, by gaining mass from a companion \citep{mccrea}, or from a direct collision \citep{hills}, in order to remain in the location of the zero-age main sequence (ZAMS) while single MS stars of similar masses have turned off to become red giants. Other exotic sources include binary systems such as cataclysmic variables (CVs; a white dwarf (WD) star accreting mass from a companion), X-ray binaries and millisecond pulsars (MSPs, \citealt{Shara_2006, ivanova, hong}). 

CVs and low-mass X-ray binaries (LMXBs; a neutron star or black hole accreting material from a low-mass companion), are faint at optical wavelengths and not easily detectable in globular clusters as they are surpassed by the large number of MS stars that shine brightly in the optical. Instead, we search for these exotic sources in the ultraviolet and X-ray wavelengths. Since the accretion of material in these binary systems results in very hot temperatures in the accretion disk, as well as when material falls onto the surface of the compact object, the spectral energy distribution of this hot emission peaks in the far-ultraviolet ($FUV$) or X-ray wavelengths. Since the surfaces of the numerous MS and red giant branch (RGB) stars in globular clusters do not reach such high temperatures, they are faint at these wavelengths and the cluster core appears much less crowded. By taking observations in the ultraviolet, we can more easily detect and identify such exotic sources. 
           
For the first part of this investigation (\citealt{m30paper1}, hereafter \hypertarget{paper1}{Paper I}), we performed the photometry of ultraviolet observations of the globular cluster M30 (NGC 7099) and presented a $FUV - UV$ CMD and an analysis of the radial distributions of the different stellar populations. A more detailed overview of M30 and the previous studies on this cluster can be found in \hyperlink{paper1}{Paper I}. We also presented an ultraviolet catalogue of sources\footnote{\url{http://vizier.u-strasbg.fr/viz-bin/VizieR?-source=J/MNRAS/511/3785}}. In this present work we continue our investigation into the properties of M30 and compare the positions of the sources detected in the ultraviolet to the locations of the known X-ray sources in M30. In this work we also perform a variability study using the $FUV$ exposures to find active binary systems in M30, including those in the vicinity of X-ray sources.           
           
Stars can exhibit brightness variability due to the orbit of a binary (or multiple) system,  structural processes within single stars which generate pulsations, or from interactions between binary members such as mass transfer. For example, the emission from a hot accretion disk in a CV system, as well as the fall of material onto the surface of the WD, can produce strong variability which helps to distinguish CVs from isolated WDs. Investigating brightness fluctuations allows for precise stellar identification within the population groups. For instance, within M30, seven RR Lyrae variables have been observed on the horizontal branch (HB) in the optical and infrared \citep{Pietrukowicz, kains}, and since \citet{zurek} saw brightness variability on the HB in the \textit{FUV} for NGC 1851, a variability study of M30 in the \textit{FUV} will be valuable to find other potential RR Lyrae stars on the horizontal branch of M30, as well as possible CV candidates. 

M30 also contains a number of X-ray sources which indicate exotic binary systems. \citet{lugger} detected 13 X-ray sources that were also found by \citet{zhao}, who discovered an additional 10 sources, based on observations with \textit{Chandra} ACIS-S. Some of these sources are thought to be CVs, and one is likely a quiescent low-mass X-ray binary (qLMXB). Two milli-second pulsars were also found in M30 by \citet{ransom}.

The observations and coordinate transformations used for this investigation are described in Sect. \ref{observation}. The potential ultraviolet counterparts to the X-ray sources and their locations in the $FUV - UV$ CMD are described in Sect. \ref{xraysec}. The variable sources are described in Sect. \ref{variable}, followed by a summary in Sect. \ref{summ}.

\begin{figure}
    \centering
    \includegraphics[width=\columnwidth]{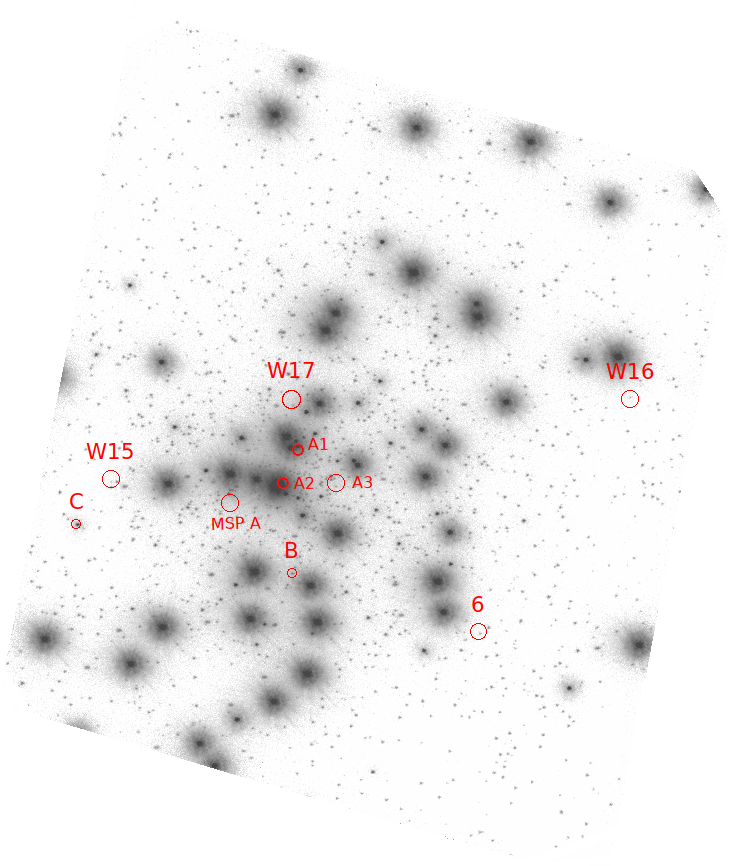}
    \caption{Positions of the 10 X-ray sources listed in Table \ref{xray}, seen as red circles with sizes proportional to their positional uncertainties,
superimposed onto the $FUV$ master image of M30. North is up, east is to the left and the image spans 34.6$''$ $\times$ 30.8$''$.}
    \label{xrayall}
\end{figure}

\section{Data processing}\label{observation}

\subsection{Photometry}

The far-ultraviolet (\textit{FUV}) data was obtained with the Advanced Camera for Surveys (ACS) on board the \textit{HST}, using the Solar Blind Channel (SBC) and the F150LP filter. The \textit{UV} data was acquired using the Wide Field Planetary Camera 2 (WFPC2) and the F300W filter, which was also on board the \textit{HST}. These were both part of the program GO-10561 (PI: Dieball) and were taken using 15 orbits distributed over three visits, from the 29th May to the 9th June 2007. The SBC has a field of view of 34.6$''$ $\times$  30.8$''$ and a pixel scale of 0.034$''$ $\times$ 0.030$''$ pixel$^{-1}$. 
Sixty-four images were taken, each with an exposure time of 315 s, resulting in a total exposure time of 20160 s. The WFPC2 consists of four cameras, the three Wide Field Cameras (WFCs) each with a field of view of $50'' \times 50''$, and the Planetary Camera (PC) with a field of view of $34'' \times 34''$, and a pixel scale of 0.046$''$ pixel$^{-1}$. As the PC was centred on the cluster core and the region observed for the $FUV$ imaging, only the exposures from this camera were used in this work. Twenty-four images were taken with a total exposure time of 12000 s. The observations, image processing and data reduction are described in detail in \hyperlink{paper1}{Paper I}. Out of 1934 sources detected in the $FUV$, 1569 have $UV$ counterparts, including 1218 MS stars, 41 horizontal branch (HB) stars, 47 blue stragglers (BS), 185 RGB stars and 78 WD/Gap objects (sources blueward of the MS and with $FUV \gtrsim$ 19 mag).

The optical data used in this work is the catalogue from \citet{guhathakurta}, the exposures of which were also taken with the WFPC2 on board the $HST$ (program GO-5324, PI: Yanny). Eight exposures were taken on 31st March 1994, two using the F336W filter with 100 s exposure time, two using the F439W filter and 40 s exposure time and four using the F555W filter with 4 s exposure times. The data was accessed from VizeiR\footnote{\url{https://vizier.cds.unistra.fr/viz-bin/VizieR?-source=J/AJ/116/1757}}.

\begin{table*}

    \hspace*{50pt}\caption{The positions and distances of the potential counterparts to the X-ray sources and their ultraviolet photometric quantities as found in this study. The radius $r$ in Column 2 is the distance of the X-ray source from the centre of the cluster, and the distance of the potential counterpart to the X-ray source is given in Column 7. The X-ray sources A1--6 are from \citet{lugger}, sources W15--W17 are from \citet{zhao}, and the MSP from \citet{ransom}.}
    
    \begin{tabular}{lrclccccccr}
    \hline
ID$_{X-ray}$&$r$\ \ \ \ \ &$P_{err}^a$&ID$_{FUV}$&RA&Dec&Distance&Err$^b$&Type&\textit{FUV}&\textit{FUV}$-$\textit{UV}\ \ \ \ \\
 &[arcsec]&[arcsec]& &[h:m:s]&[$^\circ$:$'$:$''$]&[arcsec]&&&[mag]&[mag]\ \ \ \ \ \ \ \\
 
 \hline\hline
A1&1.312&0.24&ID272&	21:40:22.16	&-23:10:45.85&	0.238&0.992& BS  &17.170\ $\pm$\ 0.004& -0.676\ $\pm$\ 0.006  \\
A2&1.064&0.24&  ID283&21:40:22.22 & -23:10:47.65&0.078&0.325&BS/RG&17.570\ $\pm$\ 0.013&-0.646\ $\pm$\ 0.016\\
&&&	ID290&21:40:22.23	&-23:10:47.71&	0.224& 0.933& BS&14.676\ $\pm$\ 0.001 &-1.612\ $\pm$\ 0.002  \\
A3&1.599&0.42& ID181& 21:40:22.03 & -23:10:47.85 & 0.134 &0.319& RG & 23.195\ $\pm$\ 0.066 & 5.677\ $\pm$\ 0.068\\
&& &ID187&21:40:22.03&-23:10:47.46&0.289& 0.688& RG  &22.129\ $\pm$\ 0.034   & 3.971\ $\pm$\ 0.037  \\
 & & & ID191 & 21:40:22.04 & -23:10:47.51 &0.336&0.800& RG & 21.549\ $\pm$\ 0.023 & 5.444\ $\pm$\ 0.024\\
&&&	ID806&	21:40:22.00&-23:10:47.88&	0.405& 0.964& MS & 23.050\ $\pm$\ 0.054  & 3.801\ $\pm$\ 0.059 \\
B&4.904&0.24& ID1647 & 21:40:22.18 & -23:10:52.17 & 0.005 & 0.021 & -- & 21.582\ $\pm$\ 0.026 & -- \hspace*{5pt}\\
C&11.483&0.24&	ID456&	21:40:22.96&-23:10:49.74&0.038&0.158& WD/Gap  & 19.748\ $\pm$\ 0.009  & -0.821\ $\pm$\ 0.019  \\
6&11.718&0.41&	ID1700 & 21:40:21.51 & -23:10:55.15 & 0.090 &0.220 & -- & 23.256\ $\pm$\ 0.048 & -- \hspace*{5pt}\\
W15&9.559&0.43&ID449&21:40:22.81&-23:10:47.61&0.292&0.679&MS&22.298\ $\pm$\ 0.029&3.464\ $\pm$\ 0.033\\
&&&ID452&21:40:22.83&-23:10:47.88&0.364&0.847&MS&22.453\ $\pm$\ 0.031&3.483\ $\pm$\ 0.035\\
&&&ID912&21:40:22.83&-23:10:47.63&0.109&0.253&WD/Gap&23.971\ $\pm$\ 0.071&2.233\ $\pm$\ 0.093\\
W16&16.672&0.44&ID1029&21:40:20.97&-23:10:43.43&0.108&0.245&RG&23.692\ $\pm$\ 0.060&4.929\ $\pm$\ 0.064\\
W17&3.843&0.46&ID1681&21:40:22.18 & -23:10:43.83 & 0.297 & 0.646 & -- & 22.648\ $\pm$\ 0.060 & -- \hspace*{5pt}\\
MSP A&3.870&0.43&ID1217&21:40:22.39&-23:10:48.35&0.399&0.928&RG&22.272\ $\pm$\ 0.054&3.859\ $\pm$\ 0.057\\
\hline

    \end{tabular}
    \\
\hspace*{-420pt}a) 95\% error circle.\\

\hspace*{-313pt}b) Characteristic uncertainty, $\textnormal{Err}\:=\:\textnormal{Distance}/P_{err}$.

    \label{xray}
\end{table*}

\subsection{X-ray coordinate transformation}

\citet{lugger} found 13 X-ray sources within the half-mass radius of M30 (1.15$'$, \citealt{harris}) using \textit{Chandra} ACIS-S. In order to check the positions of these sources against our ultraviolet dataset, we use the astrometric reference star adopted by \citet{guhathakurta}, a red giant (their source \#3611), which in our catalogue is ID366. From our dataset, the position of this star is:
\begin{equation}
\ \ \ \ \     \alpha = 21^h40^m22\fs31,\ \ \ \delta =  -23^{\circ}10'40\farcs13\ \ \ \ \textnormal{(J2000.0)}
\end{equation}

\noindent which has a shift of $\Delta\alpha$\ =\ +0.02$''$, $\Delta\delta$\ =\ +0.53$''$ from the position of $\alpha$ =\ $21^h40^m22\fs29,\ \delta = -23^{\circ}10'39\farcs60$ given by \citet{lugger} (and also a relatively small shift of $\Delta\alpha$\ =\ --0.004$''$, $\Delta\delta\ =\ $+0.03$''$ to the position of $\alpha$ =\ $21^h40^m22\fs314,\ \delta = -23^{\circ}10'40\farcs10$ reported by \citet{ransom}). This is within the uncertainties of 0.70$''$ in the world coordinate system (WCS) for both \textit{HST} and \textit{Chandra}. This positional shift is then applied to the \citet{lugger} $\textnormal{X-ray}$ sources in order to match to our dataset. These 13 sources were also detected by \citet{zhao} as well as an additional 10 sources within the updated half mass radius of 1.03$'$ \citep{harris2010}. By comparing the coordinates given for the 13 sources discovered by \citet{lugger} to those of \citet{zhao}, an additional shift can be applied to the new 10 sources to match them to our dataset. Out of the 23 total sources, 10 are within the $FUV$ field of view and are marked in Fig. \ref{xray}. After comparing the datasets, we find six confident matches: ID283 to A2, ID1647 to B, ID456 to C, ID1700 to 6, ID912 to W15 and ID1029 to W16 (described in detail in Sect. \ref{xraysec}). The average distance of these six sources to their X-ray counterparts was 0.148$''$, which is then used as an additional boresight correction to shift the X-ray catalogue into the $FUV$ frame. After this shift the confident matches have an average positional offset of 0.071$''$. 

\subsection{Brightness variability}

For the variability study, the magnitudes of the individual sixty-four $FUV$ exposures were measured using the \texttt{DAOPHOT} task \citep{stetson}, running under \texttt{PYRAF} \citep{pyraf}. As the individual images are slightly shifted with respect to one another, a specific coordinate list of the sources is first created for each exposure, using the coordinate list of the drizzled master image from \hyperlink{paper1}{Paper I}. This list is  transformed into individual coordinate lists using the task \texttt{WCSCTRAN} which takes the WCS information from each exposure. Then the \texttt{DAOPHOT} task performs aperture photometry on each exposure using the specific coordinate list. This task was performed without recentering which would measure the magnitude of a nearby bright star if the flux at the target coordinate is faint. Aperture and sky corrections to the measured fluxes are applied using the same method described in \hyperlink{paper1}{Paper I} (Sect. 2.1.2).

\section{X-ray source matching}\label{xraysec}

\begin{figure}
    \centering
    \includegraphics[width=\columnwidth]{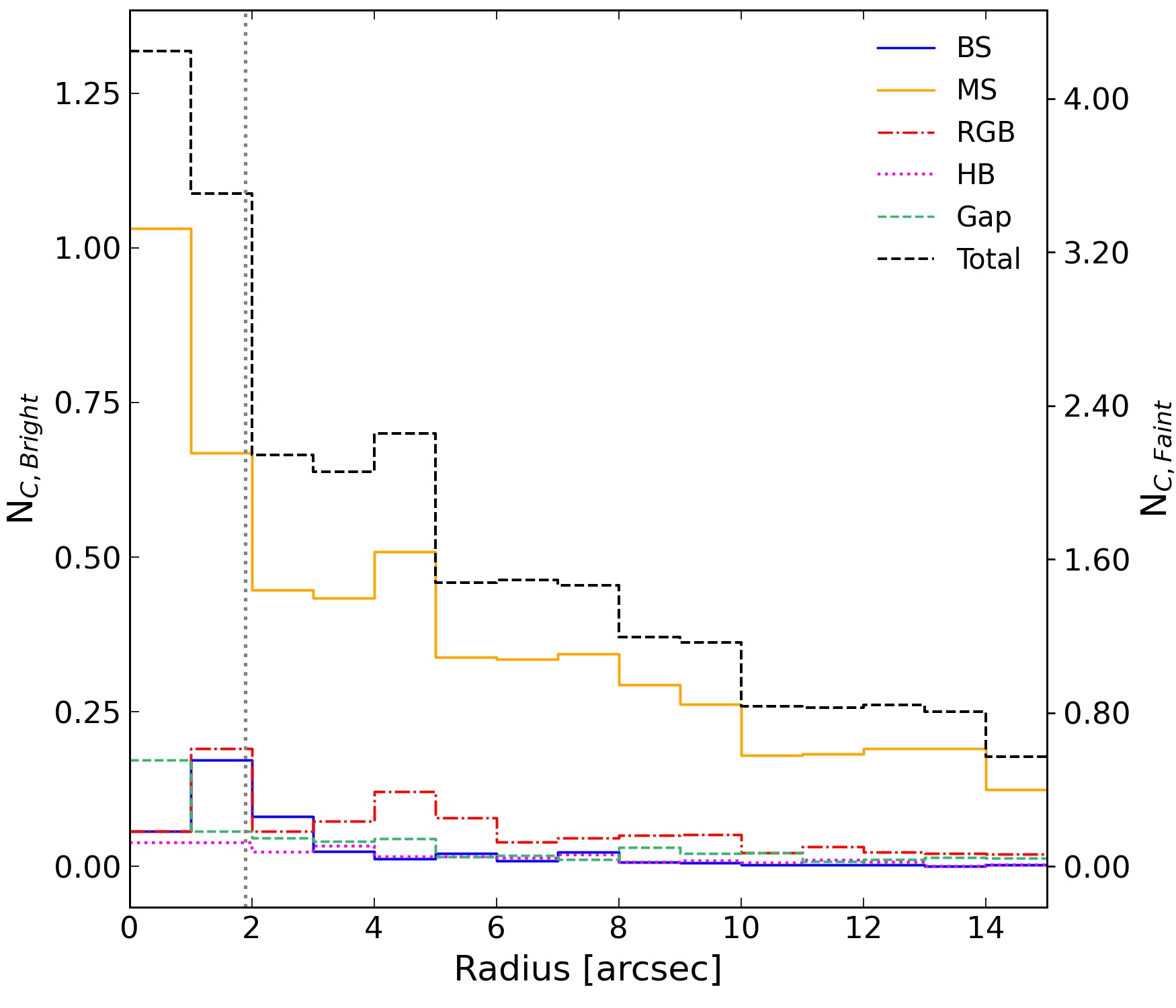}
    \caption{Number of chance coincidences within the error circle for either a bright or a faint X-ray source plotted as a function of radial distance from the centre of the cluster. For a bright source the error radius $A_{err} = 0.18$ arcsec$^2$ is used and for a faint source $A_{err} = 0.58$ arcsec$^2$ is used. The dotted vertical line indicates the cluster core region. The group Total gives the chance of finding any coincidental star in the error radii.}
    \label{chance}
\end{figure}

\subsection{Positional error radii}

\begin{figure}
    \centering
    \includegraphics[width=\columnwidth]{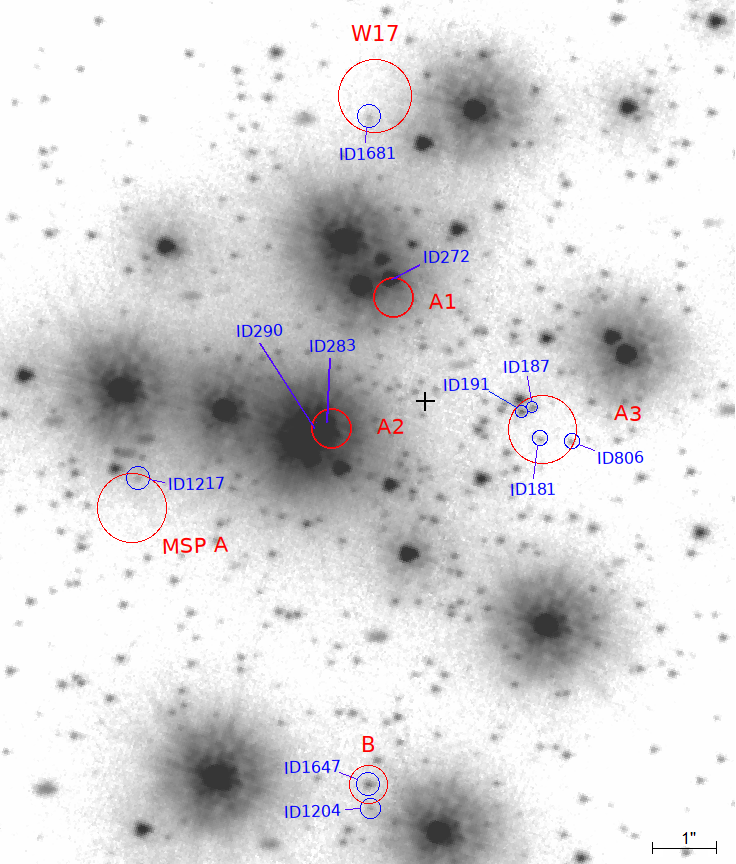}
    \caption{Position of X-ray sources with 95\% error radii marked with red circles. The potential counterparts are given in blue. The black cross indicates the centre of the cluster.}
    \label{A1B}
\end{figure}

\begin{figure}
    \centering
    \includegraphics[scale=0.507]{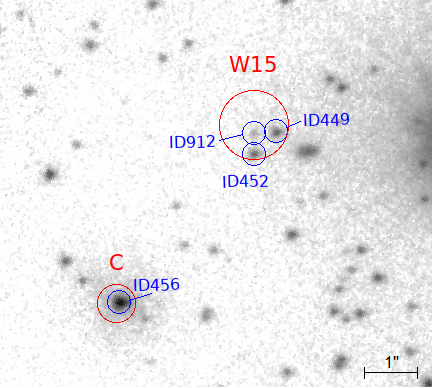}
    \includegraphics[scale=0.48]{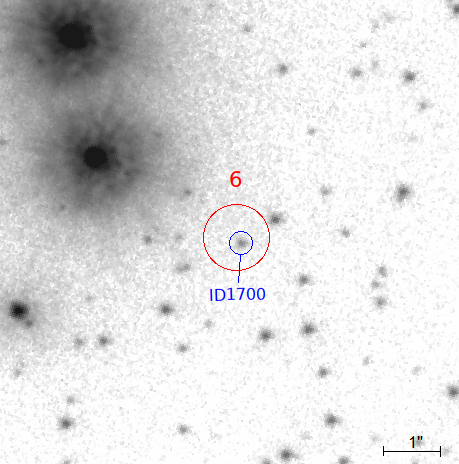}
    \includegraphics[scale=0.55]{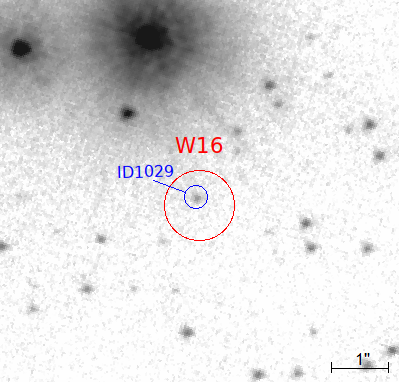}
    \caption{Same as Fig. \ref{A1B}.}
    \label{xrayimages}
\end{figure}

The X-ray sources are given in Table \ref{xray} along with their potential counterparts. The positional 95\% error radii $P_{err}$ is included for each X-ray source and is shown in Figs. \ref{xrayall}, \ref{A1B} and \ref{xrayimages} as red circles. For the bright sources (> 100 counts, i.e. sources A1, A2, B and C), we calculate $P_{err}$ using the positional uncertainties of the six confident matches and the formula:

\begin{equation}
\ \ \ \ \ \ \ \sigma_r^2 = \sigma_{RA}^2 + \sigma_{Dec}^2
\end{equation} 

\noindent After applying the shifted boresight correction, the rms errors in RA and Dec are 0.076$''$ and 0.063$''$ respectively, correlating to 1$\sigma$ in each of the two directions. A 95\% error radii corresponds to 2.45$\sigma$ and as such the resulting $P_{err}$ for the bright sources is 0.243$''$. 

For the fainter sources, we take the 95\% error radii for \textit{Chandra} ACIS-S given by \citet{zhao} and calculated from the individual source count and offset \citep{hong2005}. Then all $FUV$ sources within $P_{err}$ are marked with blue circles in Figs. \ref{A1B} \& \ref{xrayimages} and the locations of these sources in the $FUV - UV$ CMD are marked in Fig. \ref{varmatches}, which includes their optical counterpart locations in the $B - V$ CMD from the \citet{guhathakurta} catalogue. The coordinate transformation to the optical catalogue was described in \hyperlink{paper1}{Paper I}.

\subsection{Chance coincidences}

In the dense core of the cluster there may be stars within the positional error radii that are chance coincidences. To estimate the number of chance coincidences of each stellar population within these radii, we follow the method of \citet{zhao}, and split the cluster into concentric annuli with 1 arcsec thicknesses over the 15 arcsec radius contained in our field of view. Assuming the populations are evenly distributed within each annulus, we can then calculate the number of chance coincidences, $N_C$, using the number of stars in each population in each annulus, $N_{tot}$, the area of each annulus, $A_{ann}$, and the area of the 95\% positional error radii, $A_{err}$, in the formula:

\begin{equation}
    N_C = N_{tot} \times \frac{A_{err}}{A_{ann}}
\end{equation}

\noindent which we have applied to the bright and faint X-ray sources separately, using the value $A_{err} = 0.18$ arcsec$^2$ for the bright sources and the average value $A_{err} = 0.58$ arcsec$^2$ for the faint sources. The number of chance coincidences found within the error circle for either a bright or a faint X-ray source is shown for each population, as well as the total chance of finding any type of star, is given in Fig. \ref{chance}. As the area of the 95\% positional error is slightly more than three times as large for the fainter X-ray sources, the number of chance coincidences within these radii are also slightly more than triple. Within the very core of the cluster the number of a chance coincidence of any type of star is 1.32 in the area around a bright X-ray source, but drops to an average of 0.52 over the rest of the $FUV$ field of view. The average number of chance coincidences within a faint X-ray error circle is 1.66.

\subsection{Cluster membership probability}

Some of the ultraviolet sources discussed in this section and also in Sect. \ref{variable} have cluster membership probabilities calculated from their proper motions derived from the \textit{Hubble Space Telescope UV Legacy Survey of Galactic Globular Clusters} (HUGS) catalogue \citep{Piotto_2015,nardiello}. The values for these sources are given in Table \ref{member}.

\subsection{Potential X-ray counterparts}

A total of 16 $FUV$ sources are located within the 10 X-ray positional error radii. In order to see if we can already identify a number of these sources as confident matches based on their positions alone, we counted the number of $FUV$ sources within the error radii at several small and increasing offsets.  Due to the crowding in the centre of the cluster, the total number of $FUV$ sources within all 10 error radii remained roughly the same at different offset positions. This indicates that a detailed examination into the characteristics of each potential X-ray counterpart is needed to evaluate the likelihood of associating  it with the X-ray source. 

\subsubsection{Source A1}

The X-ray source A1 \citep{lugger} is thought to be a qLMXB with a neutron star that has either a He atmosphere or a H atmosphere with significant hotspots \citep{echiburu}. As LMXBs in a quiescent state have very reduced or no accretion, they are generally faint at ultraviolet and other wavelengths. \citet{zhao} note that the donor star should be a He WD if the neutron star has a He atmosphere, and it could be a MS star if the neutron star has hotspots which distort the spectrum. \citet{lugger} suggest two MS stars that lie close to the MS turnoff in the optical CMD as possible companions, however they note that the area in the vicinity of A1 is severely crowded. As A1 lies within 2$''$ of the cluster centre, the number of coincidental stars within the error radii at this radius is 1.09. From our dataset there is only one potential counterpart that lies on the edge of the error circle for A1, a BS star ID272 with a bright $FUV$ magnitude ($\approx$\;17.2 mag), making it one of the brightest BS sources, and it it also relatively bright in the optical, with $V=18.16$ mag (\citet{guhathakurta}, their source \#3115). As ID272 is a bright BS star and neither a WD or MS star, it is likely that ID272 is a chance superposition in this dense region of the cluster core, and the true counterpart is too faint to be seen in the $FUV$.

\begin{figure*}
    \centering
    
    \includegraphics[scale=0.069]{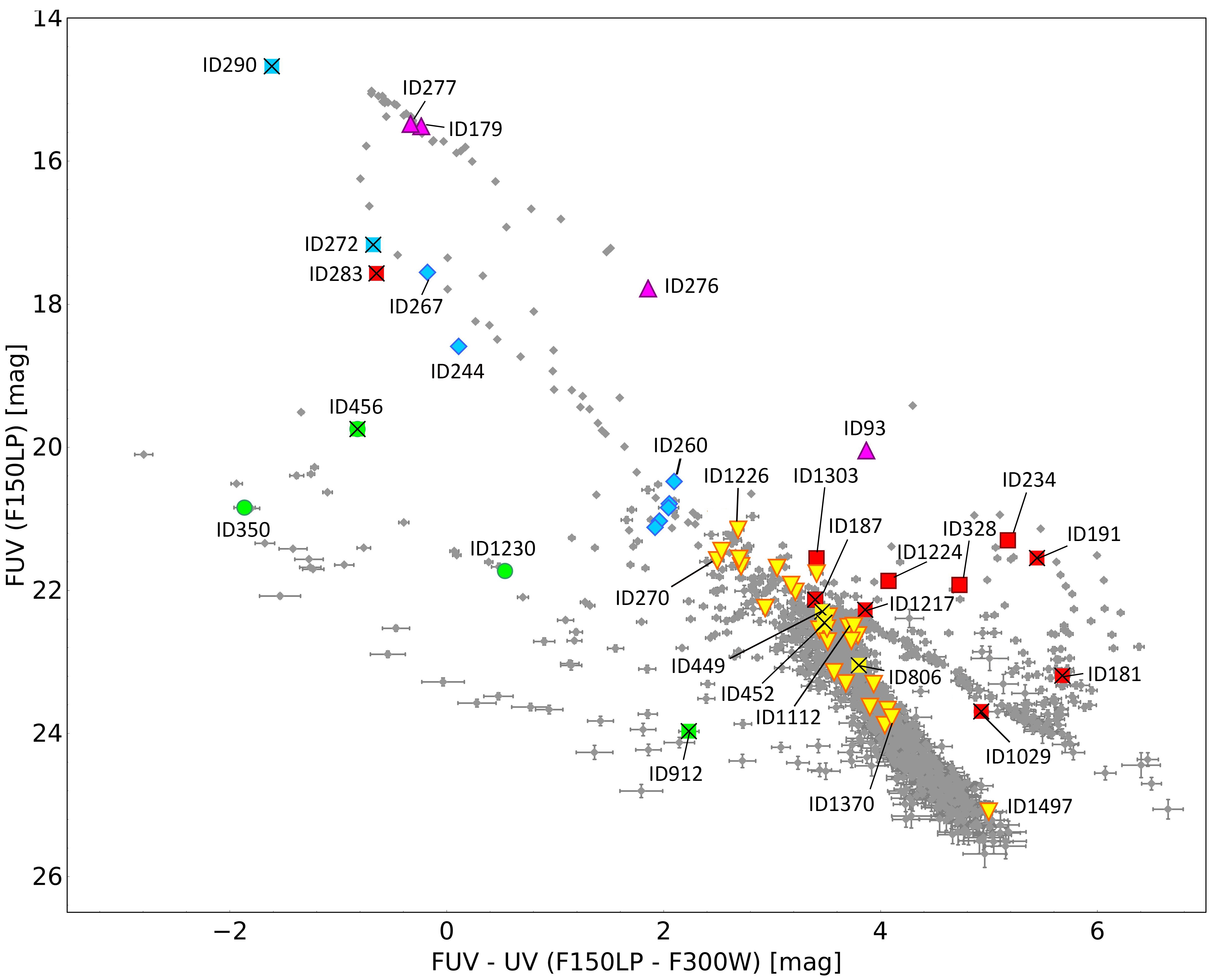}
    \hspace*{5pt}
    \includegraphics[scale=0.071]{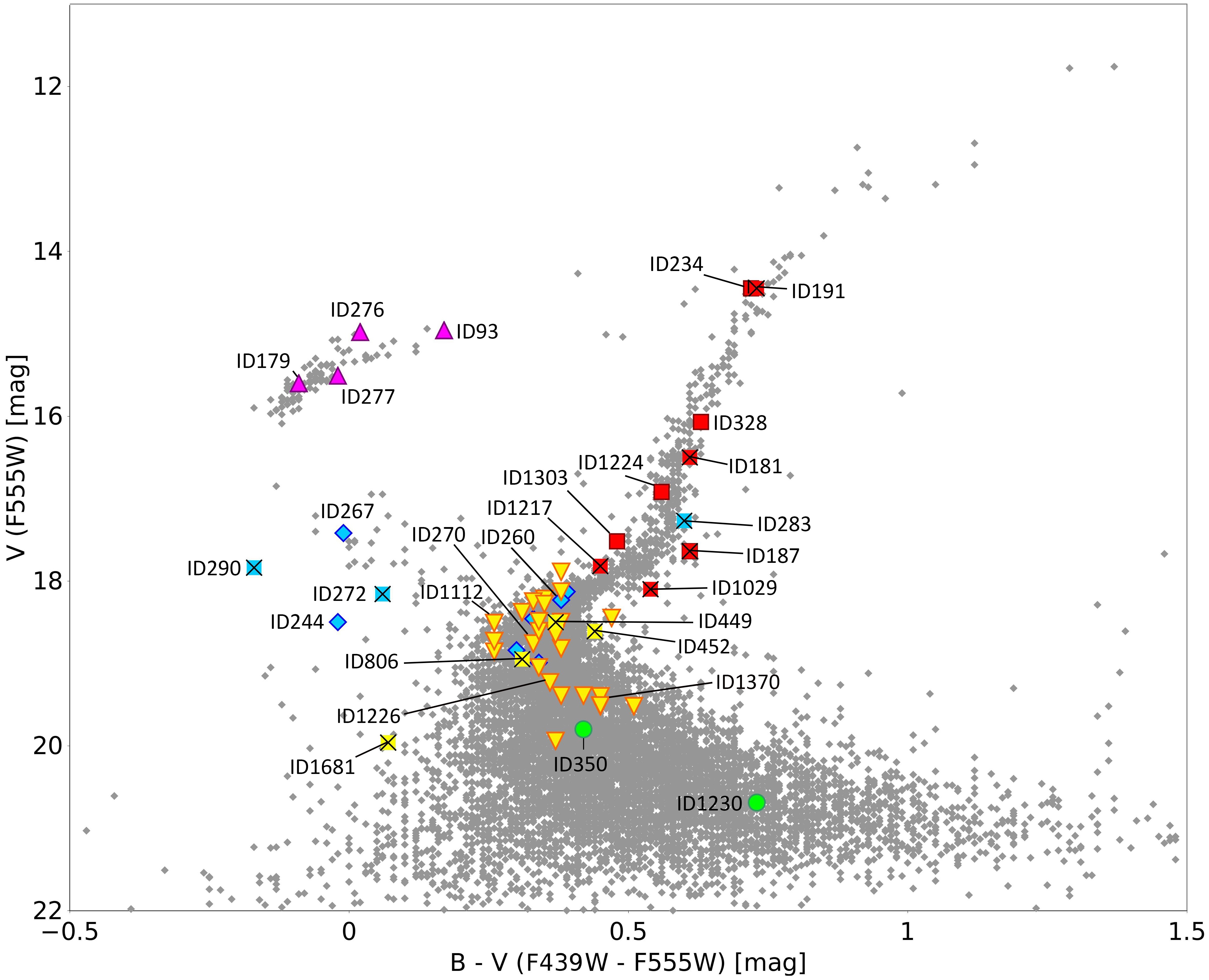}
    
    \caption{\textit{FUV} $-$ \textit{UV} (top) and optical (\citealt{guhathakurta}, bottom) CMDs with the positions of the variable BS (blue diamonds), HB (purple up triangles), MS (yellow down triangles), RGB (red squares) and WD/Gap (green circles) sources marked. The colours and shapes in one CMD are given based on the source location in the other CMD. Those with measured periods are numbered and those without can be found in Table \ref{vartable}. Additionally, the potential counterparts to the X-ray sources \citep{lugger,zhao} are marked with crosses and are numbered. The reason that source ID283 is marked differently in the two CMDs is discussed in the text.}
    \label{varmatches}
\end{figure*}

\subsubsection{Source A2}

\citet{lugger} consider the brightest \textit{FUV} source (\#3327 in \citealt{guhathakurta}, ID290 in this work) as a possible companion to the X-ray source A2 as it sits within 0.50$''$ of it.  However they note that the crowding in this central region of the cluster (Fig. \ref{A1B}, $N_{C,Bright}$ = 1.09) may mean that a different star may be the companion. From our dataset, the error circle for A2 includes source ID290 at a distance of 0.224$''$ which is located in the BS region in the optical CMD (Fig. \ref{varmatches}), but has the brightest \textit{FUV} magnitude in this work. \citet{lugger} conclude that this source is likely to be a sub-dwarf B star (sdB) with a WD companion. Yet, as noted by \citet{lugger}, not many sdB binaries are known to emit X-ray emission and thus its proximity to A2 could be a chance superposition. There is one other source within the error circle: ID283 at a distance of 0.078$''$ from A2. This is an interesting counterpart candidate as in the $FUV - UV$ CMD it has the position of a bright BS star, however in the optical CMD it is located on the red giant branch. This likely represents a binary system in which the RGB star has overfilled its Roche Lobe and is transferring mass to a compact companion resulting in strong $FUV$ emission. CVs undergoing substantial mass transfer appear considerably bluer in the ultraviolet relative to their optical colours. These objects are rare although have been found in other globular clusters (eg. \citealt{edmonds}). Thus it is likely then that ID283 is the counterpart to the X-ray source A2 and is a CV system with a red giant donor star.

\subsubsection{Source A3}

The source A3 was a faint X-ray detection with no optical counterparts proposed by \citet{lugger}. In our work there are four sources within the error circle of A3, all with optical counterparts.
One MS star lies on the edge of the error circle, ID806, which is located in the MS in both optical and ultraviolet CMDs. As this source does not show significant variability (Sect. \ref{variable}), this is unlikely to be the counterpart to A3. The number of chance coincidences within the faint X-ray source error radii at this distance from the centre is 1.60 for a MS star and 0.25 for a red giant. Three RGB stars are also in the error circle, making one of these likely to be the correct counterpart. The closest to the X-ray source is ID181 at a distance of 0.134$''$, and the other two ID187 and ID191 are at distances of 0.289$''$ and 0.336$''$ respectively. ID191 shows no variability, however ID181 and ID187 are interesting possibilities as counterparts to A3 as they are both slightly variable stars which could result from instabilities in accretion if they are binary systems. As seen later in Sect. \ref{variable}, ID187 has a larger standard deviation relative to sources of similar $FUV$ magnitude than ID181, and it appears bluer in the ultraviolet CMD relative to the other red giants than the optical CMD. Thus the most likely counterpart to A3 is the variable RGB star ID187, with the closer RGB star ID181 also a possibility.

\begin{table}
    \centering
    \caption{Cluster membership probabilities of some of the sources discussed in the text, from the HUGS catalogue \citep{Piotto_2015,nardiello}}
    \begin{tabular}{cc}
    \hline
     ID & $P_{\mu}(\%)$\\ \hline\hline
    93 & 92.6\\
    179 & 44.2\\
    181 & 90.2\\
    187 & 88.5\\
    191 & 88.5\\
    268 & 97.5\\
    272 & 97.1\\
    276 & 98.0\\
    277 & 96.9\\
    350 & 85.1\\
    456 & 94.3\\
    912 & 96.7\\
    1029 & 97.5\\
    1217 & 95.6\\
    1226 & 96.7\\
    1230 & 96.7\\ 
    1647 & 90.5\\
    1681 & 90.5\\
    \hline
    \end{tabular}

    \label{member}
\end{table}

\subsubsection{Source B}

\citet{lugger} suggest a faint companion to X-ray source B that lies in the halo of a RGB star and is located on the MS in their infrared CMD, but has blue excess in the ultraviolet and concluded that source B is a potential CV. However no MS star was detected at the location of source B in this present work, and the RGB star ID1204 is located just outside of the error circle for B, at a distance of 0.288$''$ and did not show any significant variability. Only one source is located within the error circle for source B, ID1647, which is only detected in the $FUV$ and has no $UV$ counterpart. At a distance of 0.005$''$ from source B, this is the likely counterpart to B and the source which showed blue excess in \citet{lugger}.  \citet{Piotto_2015} and \citet{nardiello} also find a source at this position with a cluster membership probability of 90.5\%. The light curve for ID1647 is given in Fig. \ref{fuvonly} which shows little variability, and as the detection limit is $UV = 23.6$ (\hyperlink{paper1}{Paper I}), it is likely that the companion to source B is a faint MS star, as suggested by \citet{lugger}.

\subsubsection{Source C}

One ultraviolet source is detected within the error circle of X-ray source C (Fig. \ref{xrayimages}), a WD/Gap object ID456 at a distance of 0.038$''$. This source was found by \citet{lugger} to be a CV as it was fainter in the year 1999 by 1 mag than in 1994. ID456 is shown as a variable WD/Gap object in Sect. \ref{variable} where its magnitude in the first observing epoch was $\approx$ 1.5 mag brighter than in the latter two epochs (Fig. \ref{456}). We agree then that ID456 is a CV that is experiencing a dwarf nova event. ID456 is one of the brightest Gap sources, with $FUV \approx$ 19.7 mag, however as no source was detected in the optical at this location, and the optical detection limit is $V\approx 22$, this suggests a very faint low mass MS companion. \citet{gottgens} also classify this source (ACS ID 23423) as a CV due to its broad H$\alpha$ and H$\beta$ emission detected with the Multi Unit
Spectroscopic Explorer (MUSE, \citealt{bacon})

\subsubsection{Source 6}

One source, ID1700, lies at a distance of 0.090$''$ from X-ray source 6 from \citet{lugger} and is detected in the $FUV$ but is not seen in the $UV$. Due to the detection limit of $UV = 23.6$ (\hyperlink{paper1}{Paper I}), the companion is likely a very faint MS star. 

\subsubsection{Source W15}

Of the 10 additional X-ray sources detected by \citet{zhao}, three are in our \textit{FUV} field of view: W15, W16 and W17. Three ultraviolet sources are detected within the error circle of W15: two MS stars, ID449 and ID452, and a gap source, ID912. The number of chance coincidences of MS stars in the error radius for W15 is 0.845 but for a gap source is only 0.068, meaning that the gap source is the more likely true counterpart. The gap object is also the closest to W15 at a distance of 0.109$''$ and corresponds in position to the optical counterpart for W15 proposed by \citet{zhao}, who confirmed that it is a cluster member with a probability of 96.7\% \citep{Piotto_2015,nardiello} and that this source has a blue excess in ultraviolet to infrared wavelengths, thus concluding that W15 is a CV. This source was not in the \citet{guhathakurta} catalogue, but due to its location blueward of the MS in the \textit{FUV\;--\;UV}\;CMD and its proximity to W15, we agree that ID912 is most likely a CV. 

\subsubsection{Source W16}

\citet{zhao} suggest a sub-subgiant as a companion to W16 which correlates to the position of ID1029, a source that is located just below the red giant branch in both the \textit{FUV -- UV} and optical CMDs in this work. At a distance of 0.108$''$ from W16, it is the only source detected in the $FUV$ to be within the error circle of X-ray source W16. As \citet{gottgens} find variable H$\alpha$ emission from this source using MUSE, \citet{zhao} propose that this is an RS CVn type of AB. The $FUV$ light curve for ID1029 illustrated in Sect. \ref{variable} shows that it is not more variable relative to other sources with similar $FUV$ magnitudes (Fig. \ref{stddev}).

\subsubsection{Source W17}

 Only one source was found within the error circle for W17, ID1681, another source with an $FUV$ detection but no counterpart in the $UV$, however it does have a counterpart from the optical catalogue (\citealt{guhathakurta}, their source \#3175); a star located blueward of the MS in the optical CMD. \citet{Piotto_2015} and \citet{nardiello} also find a source at this position with a cluster membership probability of 98.1\%. \citet{zhao} proposed two potential counterparts to W17, both MS stars which exhibited brightness variability, neither of which match to the position of ID1681. The light curve for ID1681 is shown in Fig. \ref{fuvonly}, which only shows a slight variation in magnitude. 

\subsubsection{MSP A}

Two MSPs were found in M30 by \citet{ransom} using radio pulsar timing, one of which (PSR J2140--2310A, MSP A) was also faintly observed in X-ray wavelengths by \citet{ransom} and is detected again by \citet{zhao} who measure a higher number of X-ray counts for MSP A. \citet{ransom} suggest a MS star companion within 0.09$''$ of this pulsar that was seen in $V_{555}$ images but not $U_{336}$ or $I_{814}$ images. This companion is not seen in our ultraviolet exposures, instead the only source within the error circle is a red giant, ID1217, located just off the MS turnoff, and at 0.399$''$ away from the position of MSP A, is likely to be a chance superposition.

\section{Variable $FUV$ sources}\label{variable}

\begin{figure}
    \centering
    \vspace*{-2pt}
    \includegraphics[width=\columnwidth]{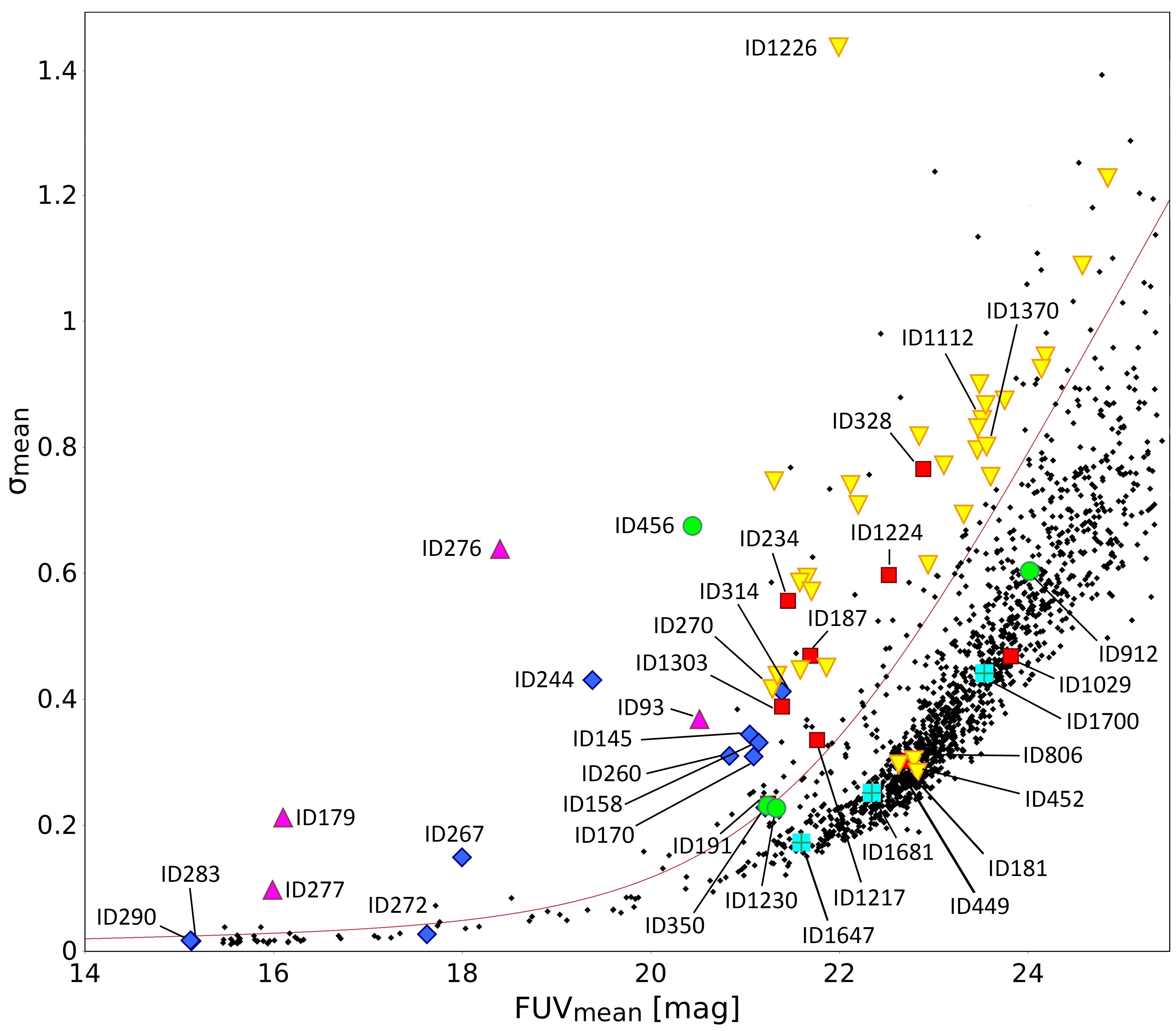}
    \caption{The standard deviation over time plotted against the mean \textit{FUV} magnitude. Marked are the variable BS sources (blue diamonds), HB stars (purple up triangles), RGB stars (red squares), MS (orange down triangles), WD/Gap objects (green circles), and sources without $UV$ measurements (blue squares). The light curves for these sources are presented in Figs. \ref{bshbrgvar} and \ref{ms1}. The faint red line indicates twenty percent above the binned average $\sigma_{mean}$.}
    \label{stddev}
\end{figure}

\begin{figure*}
    \centering
    \includegraphics[scale=0.39]{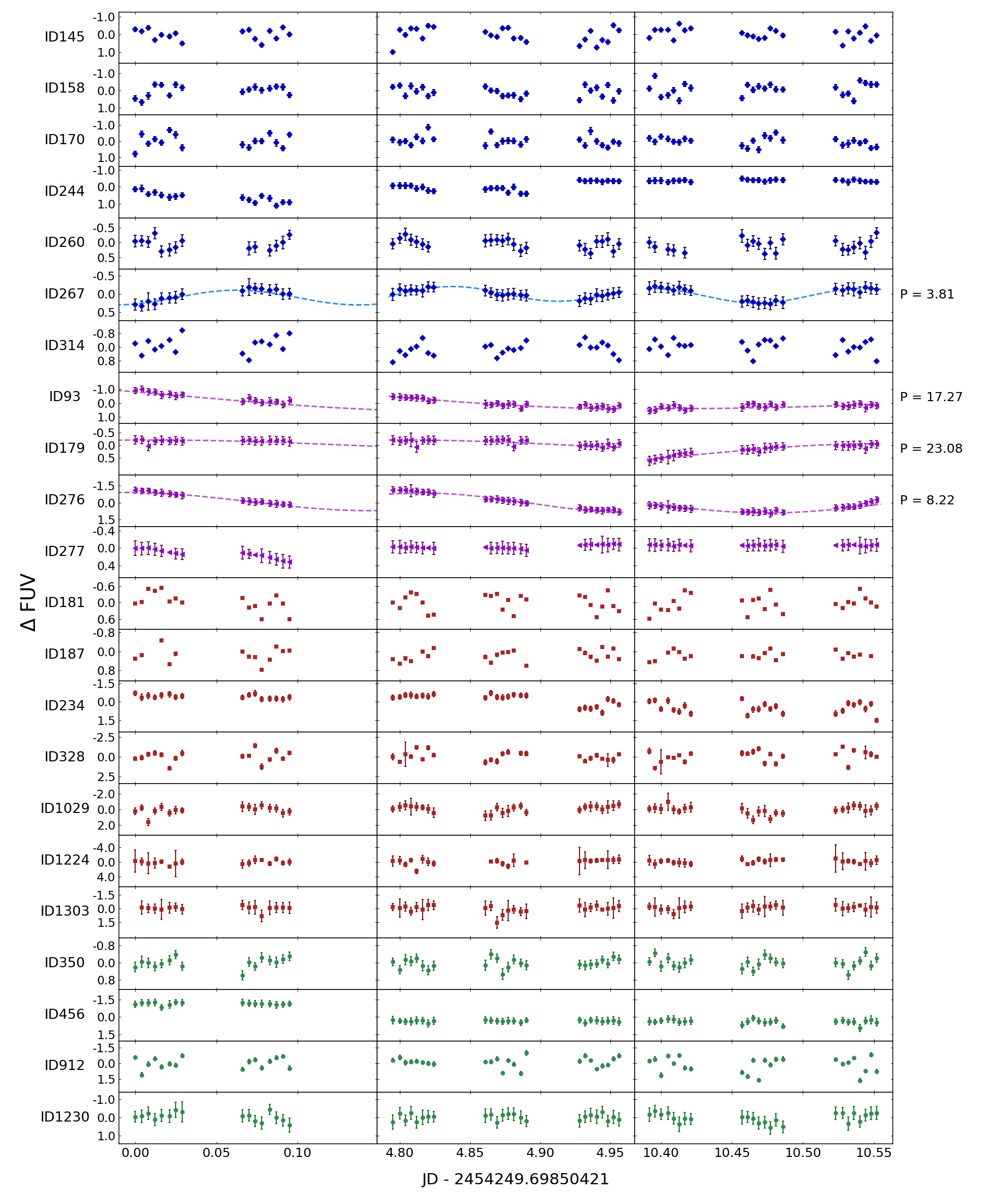}
        \caption{\textit{FUV} mean-subtracted magnitudes ($\Delta$\textit{FUV} = \textit{FUV -- FUV}$_{mean}$) are plotted over time for seven variable BS sources (blue diamonds), four HB stars (ID93 - ID277, purple triangles), seven RGB stars (red squares) and four gap objects (green circles). The sine curves represent the periods which are found for several sources, and the period in hours is given. The measured period of 3.81 hours for ID267 is half of the true period for this contact binary (see text).}
    \label{bshbrgvar}
\end{figure*}

In order to find objects in the $FUV$ dataset that exhibit significant variability, we plot the standard deviation $\sigma_{mean}$ of the $FUV$ magnitude over the mean $FUV$ magnitude for each star (Fig. \ref{stddev}). Variable sources have a larger deviation than other sources with the same magnitude and the red line in the figure indicates twenty percent above the binned average $\sigma_{mean}$ as a rough criterion for variability. From this, seven BS, four HB, five RGB and one WD/Gap source are identified as variable, and their light curves are plotted in Fig. \ref{bshbrgvar}. The two WD/Gap sources ID350 and ID1230 are also included as they are close to the variability criterion. Also included in Figs. \ref{stddev} \& \ref{bshbrgvar} are the Gap source ID912 and the RGB stars ID181 and ID1029 as these are potential X-ray counterparts. The light curves for the three sources within X-ray source error circles that are detected in the $FUV$ but with no $UV$ counterparts are given in Fig. \ref{fuvonly}. The light curves for 28 MS stars is shown in Fig. \ref{ms1} in the Appendix. The other sources above the variability criterion that are not identified were unremarkable in their light curves. The cluster membership probabilities for some of the sources discussed are given in Table \ref{member} \citep{Piotto_2015,nardiello}.

The periods of variable objects can be estimated by producing Lomb--Scargle periodograms \citep{lomb,scargle} which performs a Fourier transform of the light curve data points. In order to check for discrepancies in the data due to the observing window used, the window function is also calculated from a Lomb--Scargle periodogram with a non-floating mean model and without centering the data \citep{vanderplas}. The window function is plotted over the periodograms in Fig. \ref{periodograms} in the Appendix. The accuracy of this period is checked by comparing to the folded light curves. Four of our variable sources have periods measured and these are then used to plot sine curves representing this periodicity onto the light curves. The photometric properties and any found periods are given in Table \ref{vartable} in the Appendix for all variable sources. The uncertainties in the $FUV_{mean}$ magnitudes are larger than those for the $FUV-UV$ measurements since they are taken from the individual flat fielded exposures rather than the cleaner drizzled master image.

\begin{figure*}
    \hspace*{-30pt}
    \includegraphics[scale=0.102]{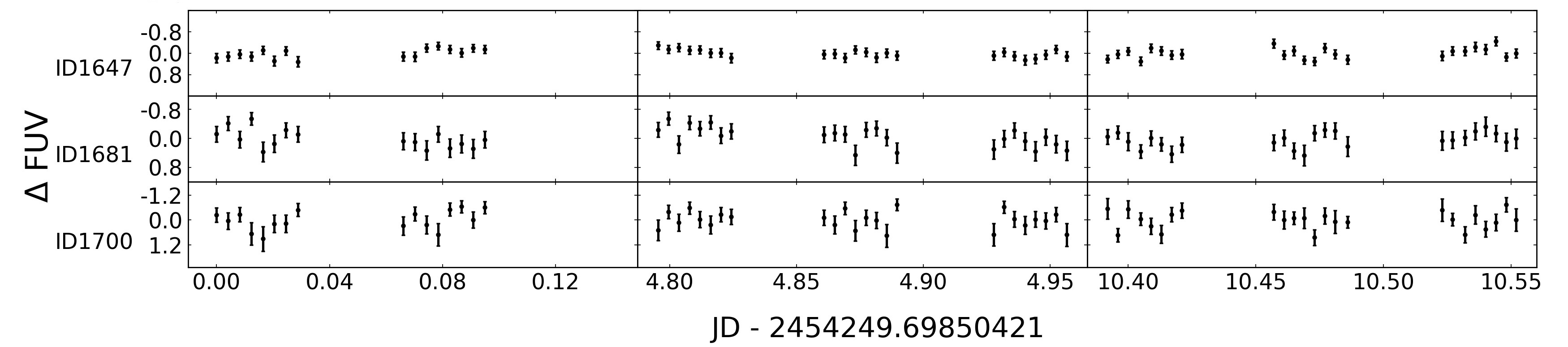}
    \caption{$FUV$ light curves for the three potential X-ray counterparts that have no matching $UV$ measurement. The mean-subtracted magnitude is plotted over time.}
    \label{fuvonly}
\end{figure*}

\begin{figure*}
    \centering

    \includegraphics[scale=0.4]{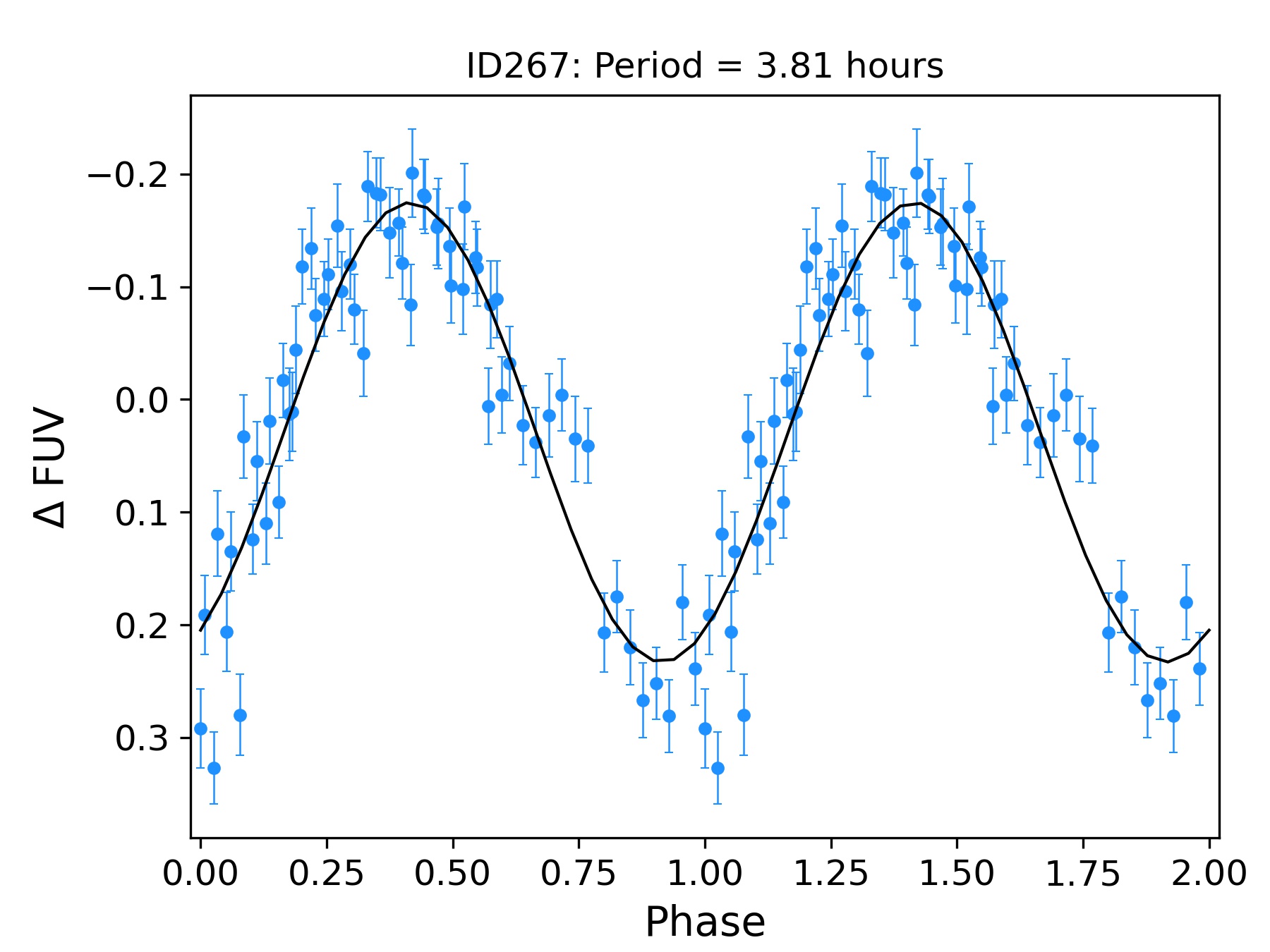} 
    \includegraphics[scale=0.4]{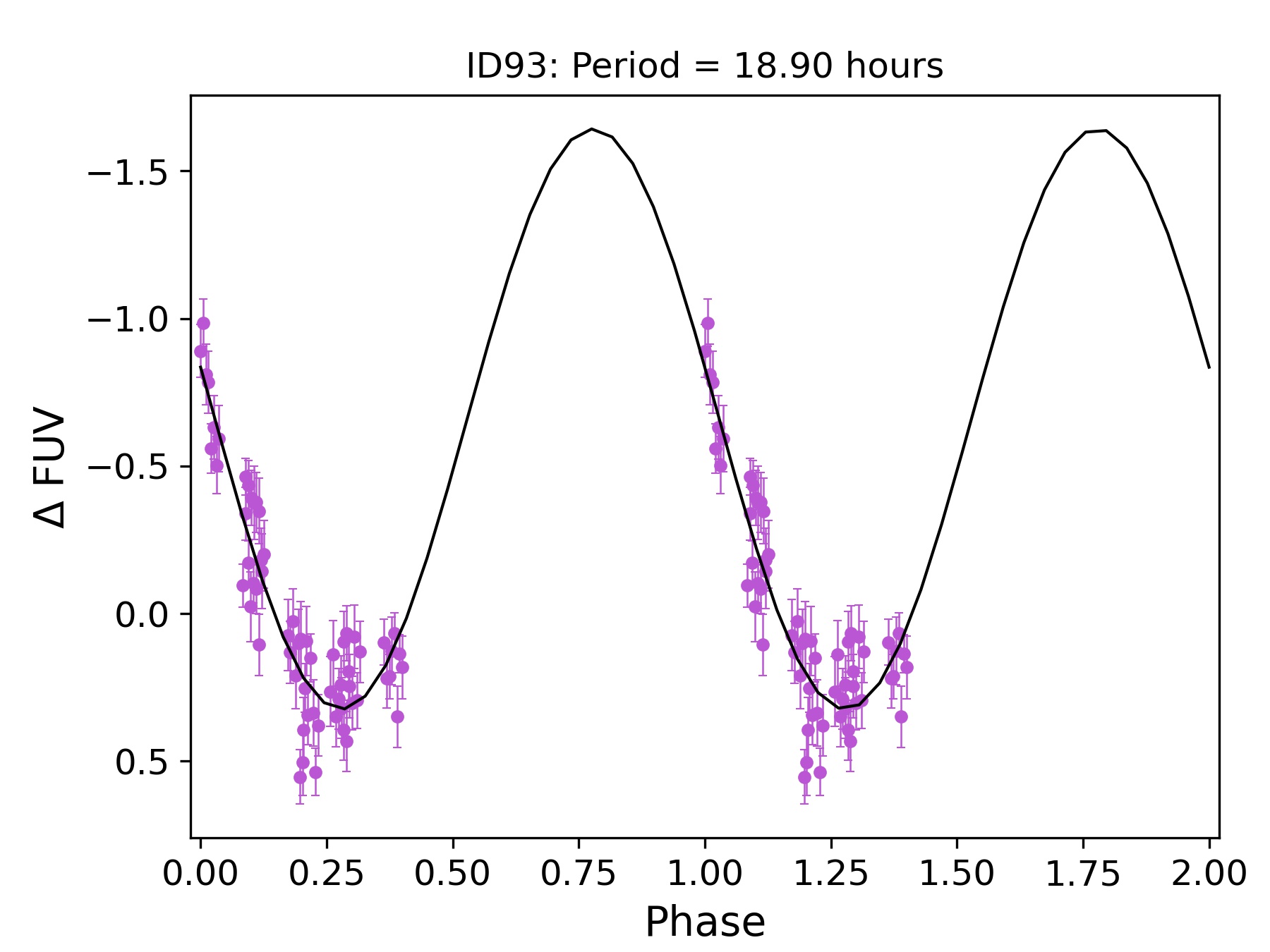}

      \includegraphics[scale=0.4]{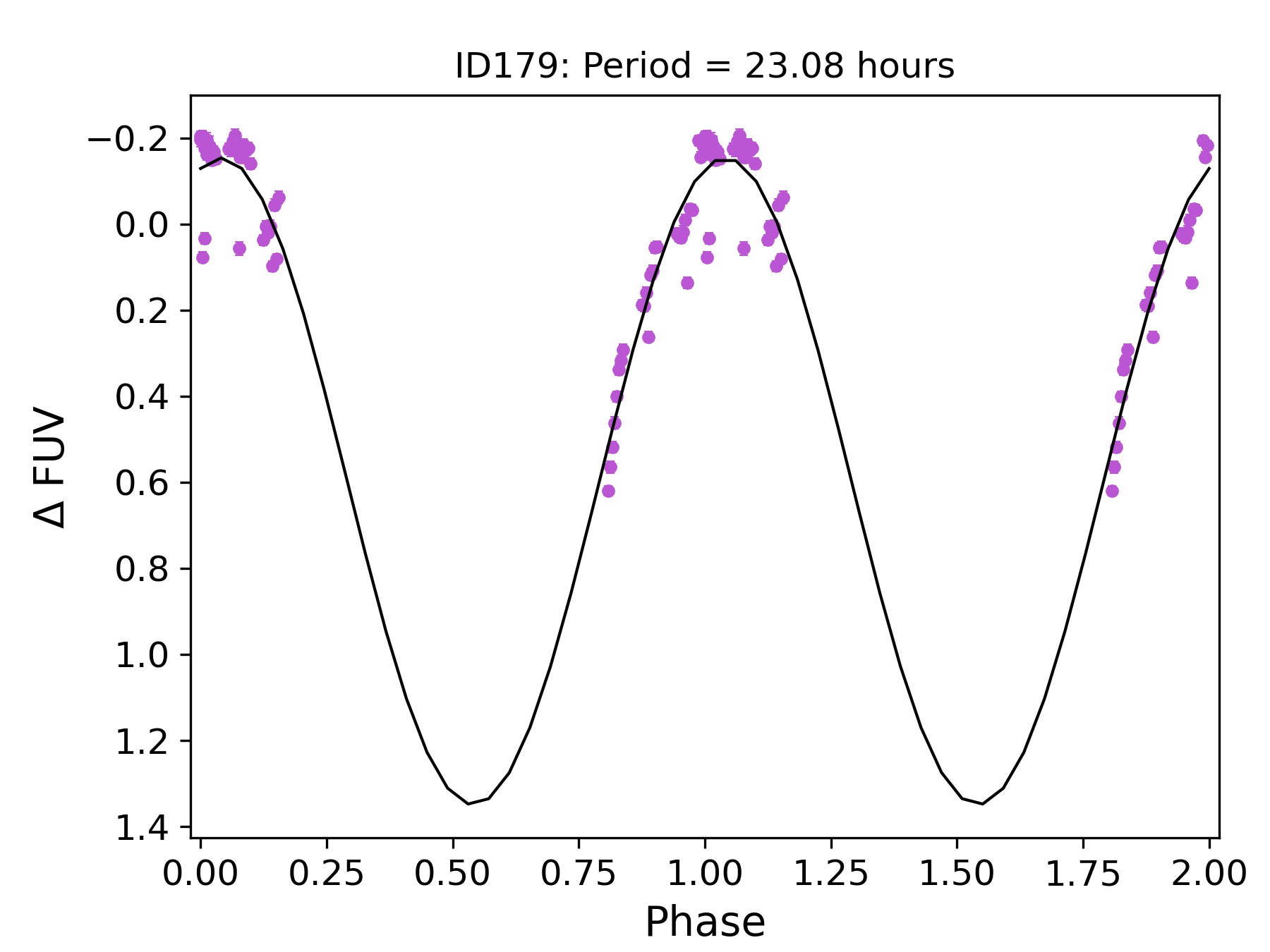}
    \includegraphics[scale=0.4]{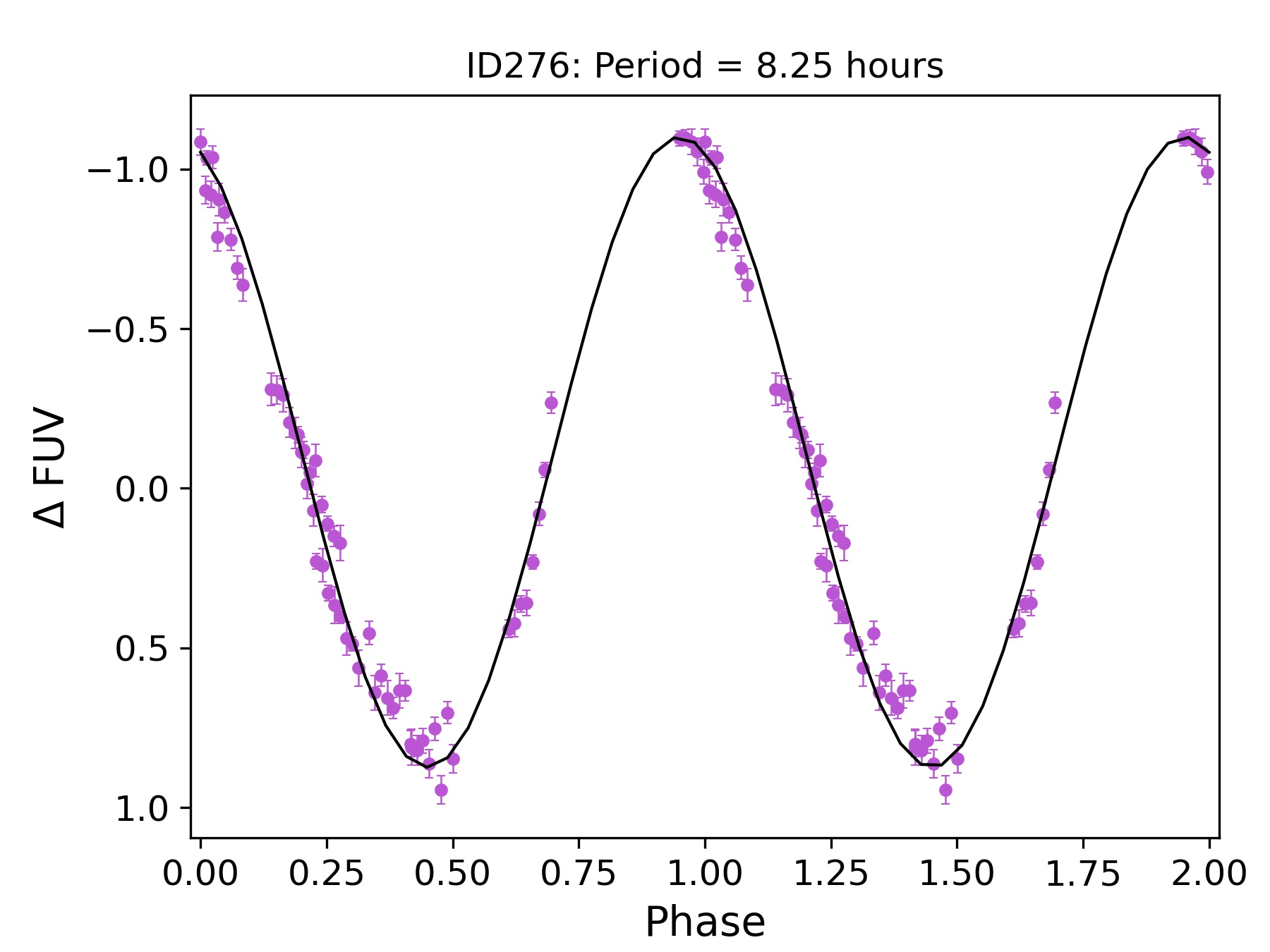}

    \caption{Folded light curves for the BS and HB stars with estimated periods shown in Fig. \ref{bshbrgvar}. The measured period of 3.81 hours for ID267 is half of the true period for this contact binary (see text)}.
    \label{periodograms}
\end{figure*}

\subsection{Blue stragglers}

The light curves of the BS stars are shown in Fig. \ref{bshbrgvar}. The source ID267 corresponds within 0.07$''$ to the position of the variable M30\texttt{\char`_}5 identified as a W UMa-type contact binary by \citet{Pietrukowicz} who measure a period of 7.61 hours. Due to the symmetry of such a system, the light curve would show a  double-peak, and in our dataset we measure the half-period of 3.81 hours (Figs. \ref{bshbrgvar} \& \ref{periodograms}), however 7.61 hours would be the true period for this system. \citet{Pietrukowicz} measured a maximum $V$ magnitude for this star of 17.28 mag which also agrees well to the value of 17.42 mag of the matched optical counterpart to ID267 from the \citet{guhathakurta} catalogue (their source \#3238). 

\subsection{Horizontal branch stars - RR Lyrae}

RR Lyrae are variable giant stars with spectral class A$-$F and are generally found on the horizontal branch in globular clusters. The sub-group RRab have steep and asymmetrical light curves with periods in the range of 12$-$20 hours, whereas the light curves for the sub-group RRc are more sinusoidal with shorter periods of 8$-$10 hours \citep{bailey,smith}. Four HB stars in our dataset were found to be variable, two of which have periods of 8.22 and 17.27 hours, indicating that these are RR Lyrae variables. ID93 matches in position within 0.12$''$ of M30\texttt{\char`_}2, identified by \citet{Pietrukowicz} as an RRab variable. Our period of 17.27 hours agrees well to the period of 16.54 hours found for M30\texttt{\char`_}2 \citep{Pietrukowicz}. These authors found a maximum $V$ magnitude for this star of 15.21 mag which also agrees well to the value of 14.96 mag of the matched optical counterpart to ID93 from \citet{guhathakurta} (their source \#2105). ID93 also correlates in position to V15 from \citet{kains} who find a period of 16.23 hours and $V=15.07$, as well as V15 from \citet{gottgens} who find that it exhibits variable H$\alpha$ emission which is typical for RR Lyrae stars. 

The other HB star, ID276, found in this work to have a period of 8.22 hours, matches within 0.15$''$ to M30\texttt{\char`_}3, an RR Lyrae variable of type RRc which has a period of 8.18 hours \citep{Pietrukowicz}. These authors measured $V_{max}$\:=\:15.08\:mag, in agreement to the optical counterpart \#3009 from \citet{guhathakurta} which had $V=14.98$\:mag. ID276 also corresponds to V19 from \citet{kains} who find a period of 8.24 hours. Our period of 8.22 hours was found by taking the second highest peak from the periodogram (Fig. \ref{periodograms}), which was double the frequency of the highest peak. This was done as the frequency of the highest peak did not produce a reasonable folded light curve and because the second highest peak corresponds well to the literature period values.

One of the other variable HB stars, ID179, is potentially a new RRab identification although it has a slightly long estimated period of 23.08 hours and also a low cluster member probability of 44.2\% \citep{Piotto_2015, nardiello}. No period was found for the other variable HB star ID277 but it is a cluster member with probability of 96.9\% \citep{Piotto_2015, nardiello}. Both sources are located on the HB in the \textit{FUV} -- \textit{UV} CMD in Fig \ref{varmatches}, with $\textnormal{\textit{FUV}} = 15.51$ for ID179 and $\textnormal{\textit{FUV}} = 15.48$ for ID277. ID179 is found at the position $\alpha = 21^h40^m22\fs25, \delta=-23^\circ10'58\farcs51$ and ID277 at $\alpha=21^h40^m22\fs03,\ \delta= -23^\circ10'39\farcs25$.

\subsection{Red giant branch stars}

As mentioned in Sect. \ref{xraysec}, sources ID181 and ID187 are located at a distance of 0.134$''$ and 0.289$''$ of the X-ray source A3 \citep{lugger} respectively. As ID187 has a larger $\sigma_{mean}$ relative to sources of similar $FUV$ magnitudes, and ID181 only shows variations comparable to other stars at the same magnitude, this suggests that ID187 may have overfilled its Roche Lobe and is transferring mass to a low-mass compact companion which results in X-ray and variable ultraviolet emission.

The RGB star ID234 is located at the tip of the red giant branch and shows stability in magnitude for the first and most of the second observing epochs but exhibits fluctuations in magnitude during the last epoch. This behaviour is also seen in MS stars ID268 and ID1226 in Sect. \ref{MSsec}, the cause of which is unknown. The sub-subgiant ID1029 was included in Fig. \ref{bshbrgvar} as it was the only source within the error circle of X-ray source W16, however as seen in Fig. \ref{stddev}, it is no more variable in the ultraviolet than other sources of similar magnitudes.  

\begin{figure}
    \centering
    \includegraphics[scale=2.7]{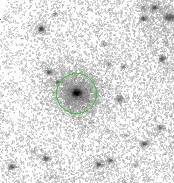} \hspace{5pt} \includegraphics[scale=2.7]{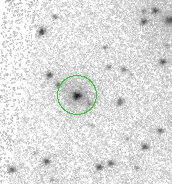}
    \caption{\textit{FUV} image of the CV candidate ID456 during dwarf nova outburst (left) and quiescence (right).}
    \label{456}
\end{figure}

\subsection{WD/Gap sources}

The \textit{FUV} light curves for four of the WD/Gap sources are shown in Fig \ref{bshbrgvar}. Source ID456 is, as indicated in Sect. \ref{xraysec}, 0.038$''$ from the X-ray source C and was proposed by \citet{lugger} to be a CV, as it had decreased in the \textit{UV} by 1 mag from 1994 to 1999. In the present work, ID456 decreases by roughly 1.5 mag in the \textit{FUV} from the first observing epoch to the second which is representative of a dwarf nova and is pictured in Fig. \ref{456}. \citet{gottgens} finds that this source exhibits broad H$\alpha$ and H$\beta$ emissions representative of accretion and it is thus very likely that this long-period CV candidate is the counterpart to the X-ray source C. 

ID912 is the likely counterpart to X-ray source W15 making this source a potential CV, which was found by \citet{zhao} to have a blue excess, however the $\sigma_{mean}$ for ID912 is not much larger than other sources of similar magnitudes.

Sources ID350 and ID1230 lie close to the variability criterion (Fig. \ref{stddev}), but their light curves only show slight variations. As such, these two sources can be considered weak variables.

\subsection{Main sequence stars}\label{MSsec}

The light curves for the MS stars which exhibit variability are given in Fig. \ref{ms1} in the Appendix. The fluctuations of many of the MS stars seem more erratic than for the other populations which may indicate starspot activity \citep{brown2011}. Some stars show little variance in magnitude over the course of one observing day, and then have large variations on another day, for example sources ID268 and ID1226 (this behaviour was also seen in the RGB star ID234). ID1226 has one of the highest $\sigma_{mean}$ values and remains stable for almost two observing epochs and then exhibits a period of large variability before settling into a stable period again.  

Source ID806 is on the edge of the error circle for X-ray source A3, and whilst it shows slight fluctuations in $FUV$ magnitude, its $\sigma_{mean}$ is no larger than sources of similar magnitude, and it remains likely that one of the RGB stars is the likely counterpart to A3.

\section{Summary} \label{summ}

We performed a comparison of the positions of the sources detected in the far-ultraviolet to known X-ray sources and possible companions or counterparts to these were discussed. The crowding in the centre of the cluster makes it difficult to determine the exact counterparts, and in this work all those that are within the 95\% error radii of the X-ray source are considered. Out of the ten X-ray sources within our field of view, six confident counterparts are: the RGB star with strong $FUV$ emission ID283 to X-ray source A2, the gap source ID456 to source C, agreeing with \citet{lugger} that this is a CV, gap source ID912 to X-ray source W15, agreeing with \citet{zhao} that this is also a CV, RGB star ID1029 to W16 which was proposed by \citet{zhao} to be a RS CVn, and the two sources detected in the $FUV$ but with no matching $UV$ counterparts: ID1647 to X-ray source B and ID1700 to source 6. Although MS stars are much more numerous in the core region of globular clusters, if we include ID187 as the possible counterpart to A3, then three of the counterparts to the ten X-ray sources are red giants.

Light curves were shown for $FUV$-variable objects and we found periods for four of these sources. Several of these variables were considered as potential X-ray counterparts. Out of the seven variable BS stars, the half-period is measured for one of them that is a known W UMa-type contact binary. Four HB stars show variability, two are known RR Lyrae variables and one (ID179) is potentially a new RR Lyrae (RRab) classification, however it may be a field star. One gap source, ID456, a previously identified CV, shows a dwarf nova event and two other gap sources have weak variability. Twenty-eight MS stars also exhibit fluctuations. 

Observations into the far-ultraviolet allow us to detect and identify exotic stellar systems such as CVs which are less easily observed at optical wavelengths amongst the numerous optically-bright MS stars. Studies using ultraviolet wavelengths and multi-wavelength comparisons continue to provide valuable insights into the different populations in dense stellar systems.

\section*{Data Availability}

The data underlying this article are available in the Mikulski Archive for Space Telescopes (MAST): \url{https://archive.stsci.edu}. The datasets are derived from images in the public domain: \url{https://archive.stsci.edu/proposal\_search.php?mission=hst&id=10561}. The catalogue of sources is available at CDS via anonymous ftp to cdsarc.u-strasbg.fr (130.79.128.5) or via the VizieR catalogue access tool \citep{vizier} at: \url{http://vizier.u-strasbg.fr/viz-bin/VizieR?-source=J/MNRAS/511/3785}. This research also made use of optical data also available at CDS: \url{https://cdsarc.cds.unistra.fr/viz-bin/cat/J/AJ/116/1757}.

\bsp	

\section*{Acknowledgements}

\vspace{5pt}

We thank the referee for the detailed and thoughtful comments.

\noindent PK acknowledges support from the Grant Agency of the Czech Republic under grant number 20-21855S.

\raggedright
\bibliographystyle{mnras}
\bibliography{M30_Mansfield2.bib} 

\appendix

\onecolumn
\renewcommand{\thefigure}{A\arabic{figure}}

\setcounter{figure}{0}

\section*{Appendix}

\captionsetup[figure]{labelfont={bf},name={Figure},labelsep=period}

\begin{minipage}[c]{\textwidth}
\centering
    \includegraphics[scale=0.4]{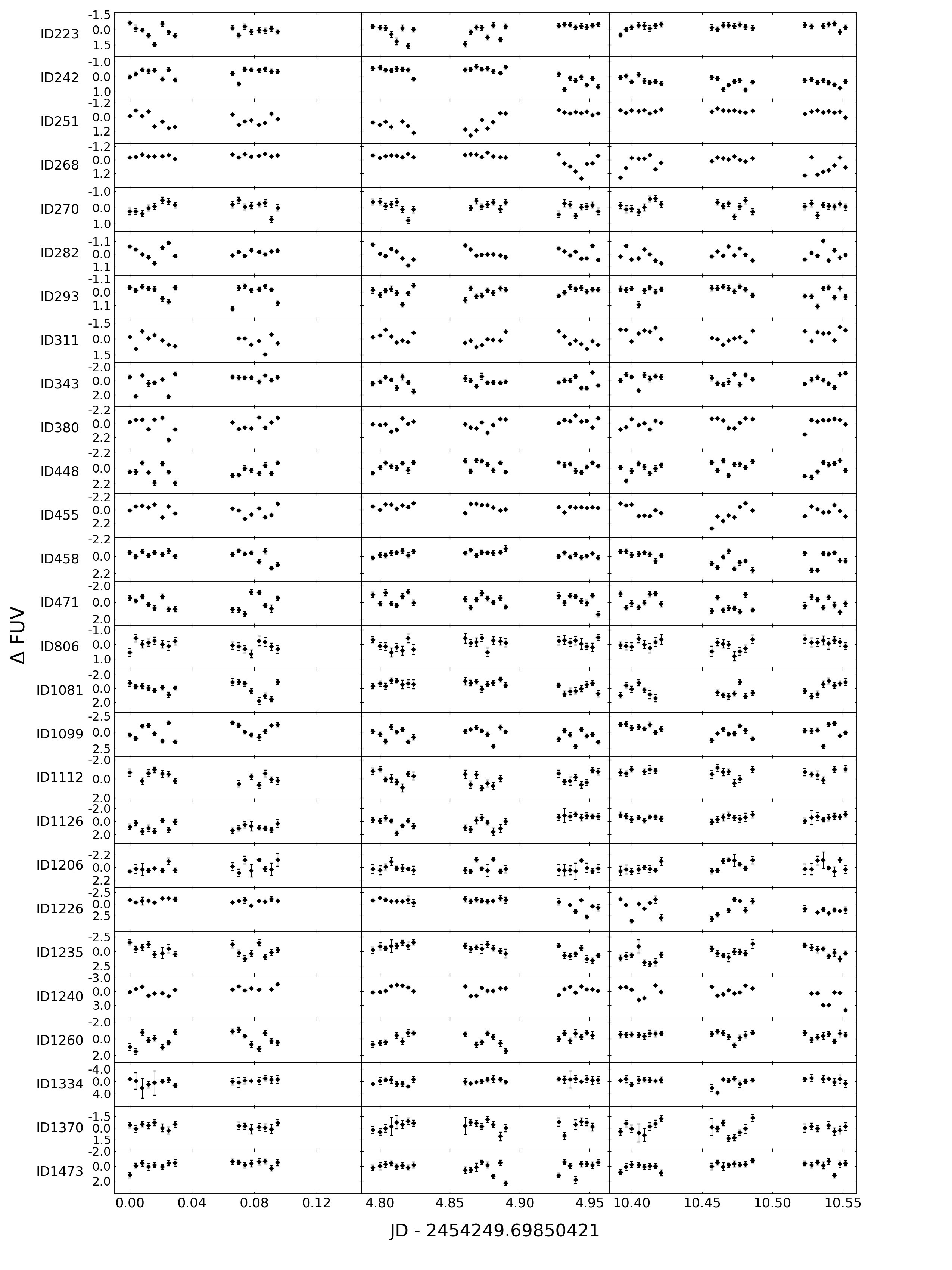}
    \captionof{figure}{Light curves for the 28 MS stars which showed variability.}
    \label{ms1}
\end{minipage}

\pagebreak
\twocolumn

\renewcommand{\thetable}{A\arabic{table}}

\begin{table}
    \centering
    
    \betweenscriptfootnote    \caption{Photometric and variability parameters of all variable sources whose light curves are shown in this work.}

    \begin{tabular}{clrcrr}
    \hline
ID&Type&\textit{FUV} - \textit{UV}\ \ \ \ &\textit{FUV}$_{mean}$\ \ \ &$\sigma_{mean}$&	Period\\
 & &[mag]\ \ \ \ \ \ \ &[mag]&&[hours]\\
 \hline\hline
93&HB&3.870\ $\pm$\ 0.011&20.518\ $\pm$\ 0.099&0.368&17.27\\
145&BS&1.961\ $\pm$\ 0.030&21.050\ $\pm$\ 0.152&0.344&\\
158&BS&1.923\ $\pm$\ 0.037&21.146\ $\pm$\ 0.161&0.331&\\
170&BS&2.054\ $\pm$\ 0.034&21.091\ $\pm$\ 0.158 &0.309&\\
179&HB&0.234\ $\pm$\ 0.002&16.103\ $\pm$\ 0.013&0.212&23.08\\
181&RG&5.677\ $\pm$\ 0.068&22.726\ $\pm$\ 0.274&0.302&\\
187&RG&3.971\ $\pm$\ 0.037&21.690\ $\pm$\ 0.172&0.469&\\
223&MS&3.215\ $\pm$\ 0.091&21.658\ $\pm$\ 0.226&0.593&\\
234&RG&5.176\ $\pm$\ 0.044&21.455\ $\pm$\ 0.196&0.556&\\
242&MS&2.688\ $\pm$\ 0.052&21.340\ $\pm$\ 0.181&0.437&\\
251&MS&2.715\ $\pm$\ 0.058&21.581\ $\pm$\ 0.221&0.585&\\
260&BS&2.098\ $\pm$\ 0.043&20.835\ $\pm$\ 0.162&0.310&\\
267&BS&0.176\ $\pm$\ 0.007&17.999\ $\pm$\ 0.034&0.149&3.81\\
268&MS&3.175\ $\pm$\ 0.050&21.703\ $\pm$\ 0.229&0.571&\\
270&MS&2.495\ $\pm$\ 0.062&21.288\ $\pm$\ 0.170&0.416&\\
276&HB&1.857\ $\pm$\ 0.005&18.401\ $\pm$\ 0.038&0.638&8.22\\
277&HB&0.332\ $\pm$\ 0.002&15.989\ $\pm$\ 0.012&0.097&\\
282&MS&2.536\ $\pm$\ 0.032&21.862\ $\pm$\ 0.278&0.450&\\
293&MS&2.697\ $\pm$\ 0.046&21.585\ $\pm$\ 0.205&0.446&\\
311&MS&3.413\ $\pm$\ 0.056&22.200\ $\pm$\ 0.403&0.708&\\
314&BS&2.046\ $\pm$\ 0.092&21.390\ $\pm$\ 0.373&0.412&\\
343&MS&3.048\ $\pm$\ 0.066&22.120\ $\pm$\ 0.357&0.740&\\
350&Gap&-1.864\ $\pm$\ 0.098&21.238\ $\pm$\ 0.169&0.231&\\
380&MS&3.522\ $\pm$\ 0.038&23.109\ $\pm$\ 0.346&0.771&\\
448&MS&3.438\ $\pm$\ 0.038&23.462\ $\pm$\ 0.409&0.795&\\
455&MS&3.793\ $\pm$\ 0.038&23.552\ $\pm$\ 0.433&0.867&\\
456&Gap&0.821\ $\pm$\ 0.019&20.443\ $\pm$\ 0.099&0.675&\\
458&MS&3.706\ $\pm$\ 0.037&23.317\ $\pm$\ 0.377&0.693&\\
471&MS&3.513\ $\pm$\ 0.041&23.603\ $\pm$\ 0.430&0.753&\\
806&MS&3.801\ $\pm$\ 0.059&22.792\ $\pm$\ 0.282&0.303&\\
912&Gap&2.233\ $\pm$\ 0.093&24.019\ $\pm$\ 0.512&0.603&\\
1029&RG&3.929\ $\pm$\ 0.064&23.817\ $\pm$\ 0.429&0.468&\\
1081&MS&3.507\ $\pm$\ 0.041&23.753\ $\pm$\ 0.473&0.874&\\
1099&MS&4.067\ $\pm$\ 0.064&24.187\ $\pm$\ 0.486&0.944&\\
1112&MS&3.756\ $\pm$\ 0.036&23.513\ $\pm$\ 0.423&0.843&\\
1126&MS&3.732\ $\pm$\ 0.047&23.469\ $\pm$\ 0.412&0.830&\\
1206&MS&2.937\ $\pm$\ 0.049&21.309\ $\pm$\ 0.149&0.746&\\
1224&RG&4.074\ $\pm$\ 0.057&22.524\ $\pm$\ 0.471&0.597&\\
1226&MS&2.491\ $\pm$\ 0.064&21.992\ $\pm$\ 0.556&1.437&\\
1230&Gap&0.539\ $\pm$\ 0.054&21.333\ $\pm$\ 0.147&0.227&\\
1235&MS&3.682\ $\pm$\ 0.055&24.143\ $\pm$\ 0.570&0.924&\\
1240&MS&4.041\ $\pm$\ 0.065&24.577\ $\pm$\ 0.712&1.088&\\
1260&MS&3.903\ $\pm$\ 0.065&22.847\ $\pm$\ 0.312&0.817&\\
1303&RG&3.412\ $\pm$\ 0.036&21.391\ $\pm$\ 0.176&0.388&\\
1334&MS&3.937\ $\pm$\ 0.053&23.484\ $\pm$\ 0.422&0.900&\\
1370&MS&4.104\ $\pm$\ 0.070&23.560\ $\pm$\ 0.433&0.801&\\
1473&MS&3.573\ $\pm$\ 0.051&22.942\ $\pm$\ 0.314&0.613&\\

    \end{tabular}
    
     \label{vartable}
\end{table}

\normalfont

\begin{figure}
    \centering
    \includegraphics[scale=0.1]{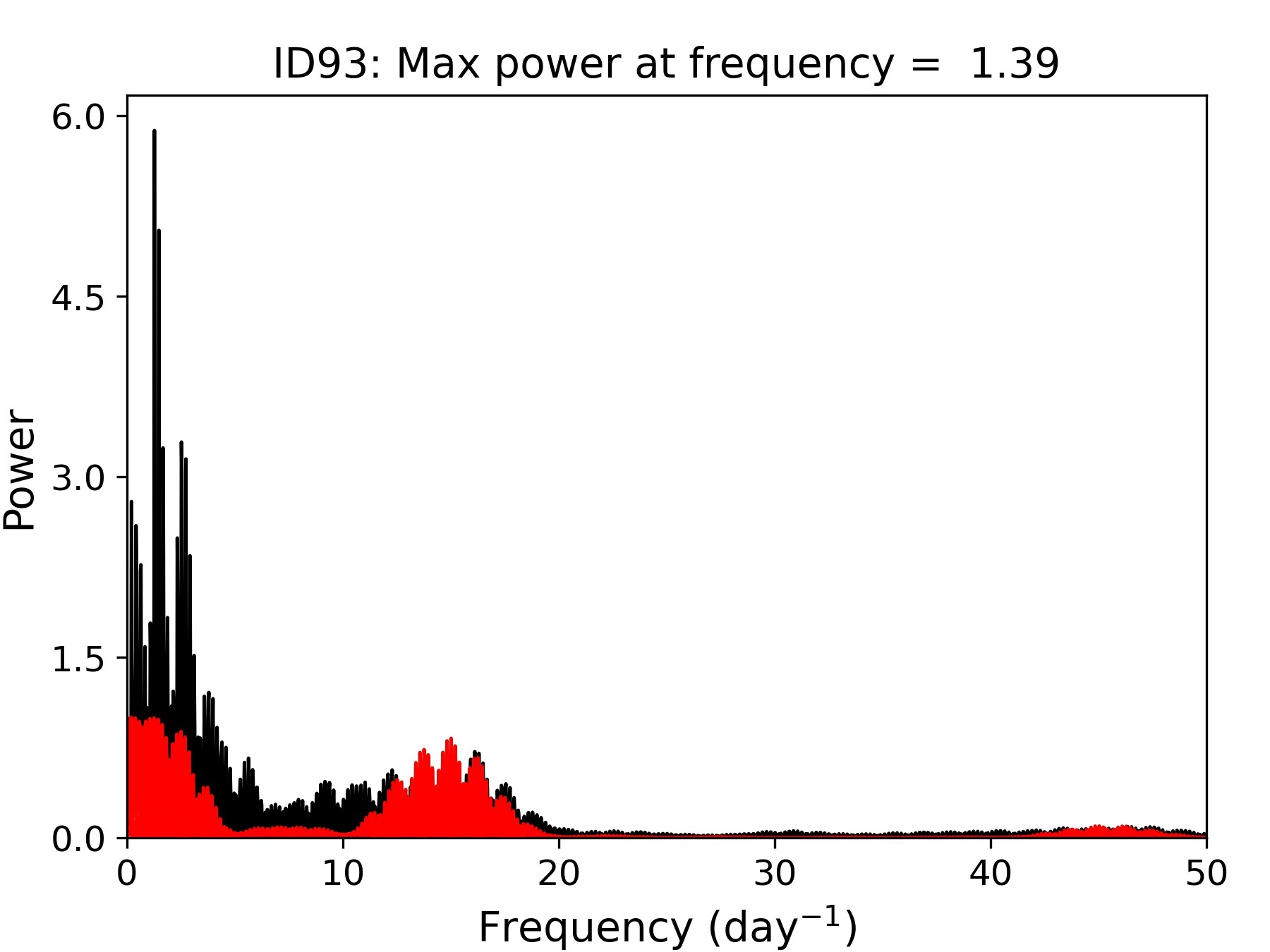}
    \includegraphics[scale=0.1]{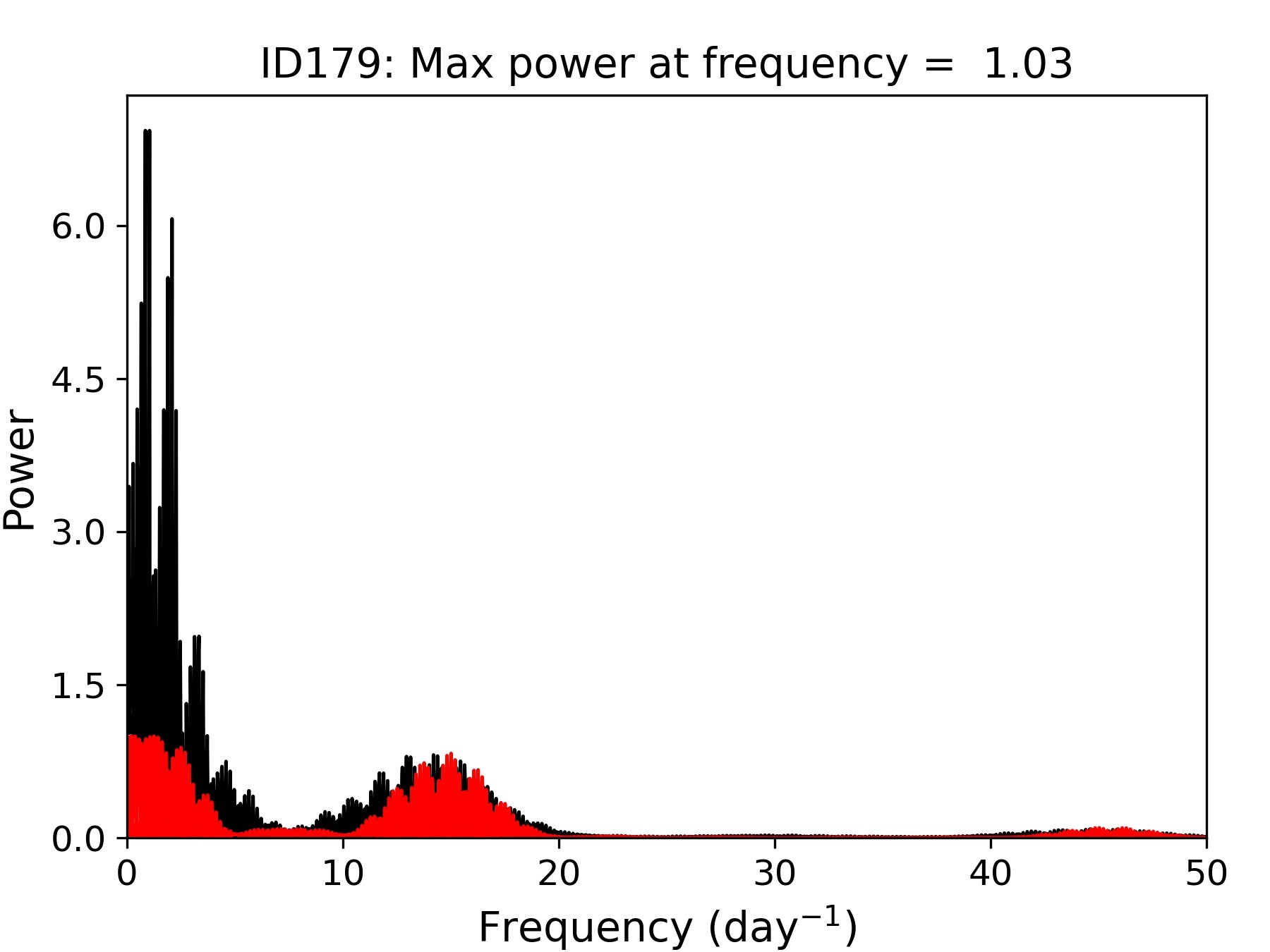}
    \includegraphics[scale=0.1]{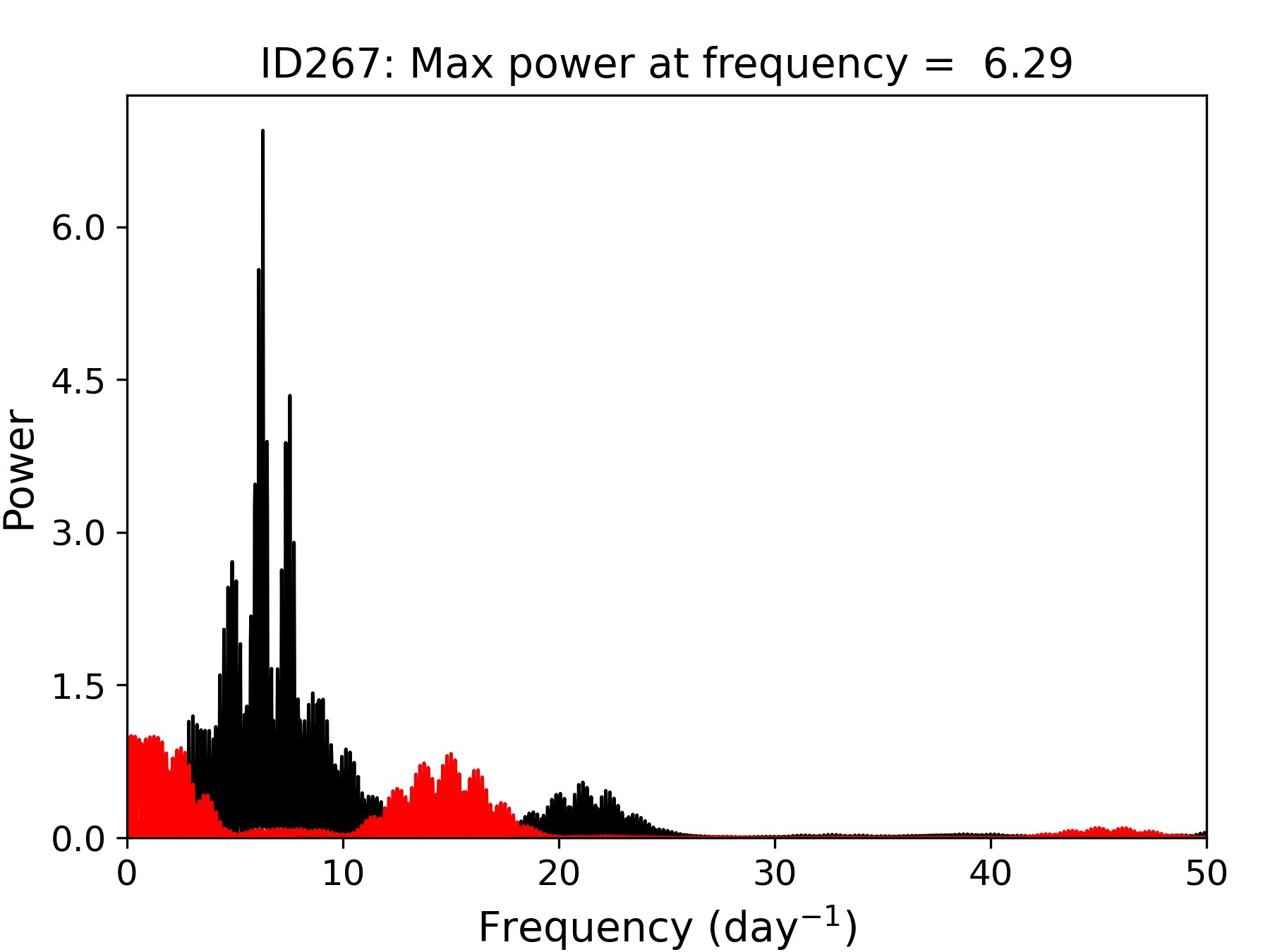}
    \includegraphics[scale=0.1]{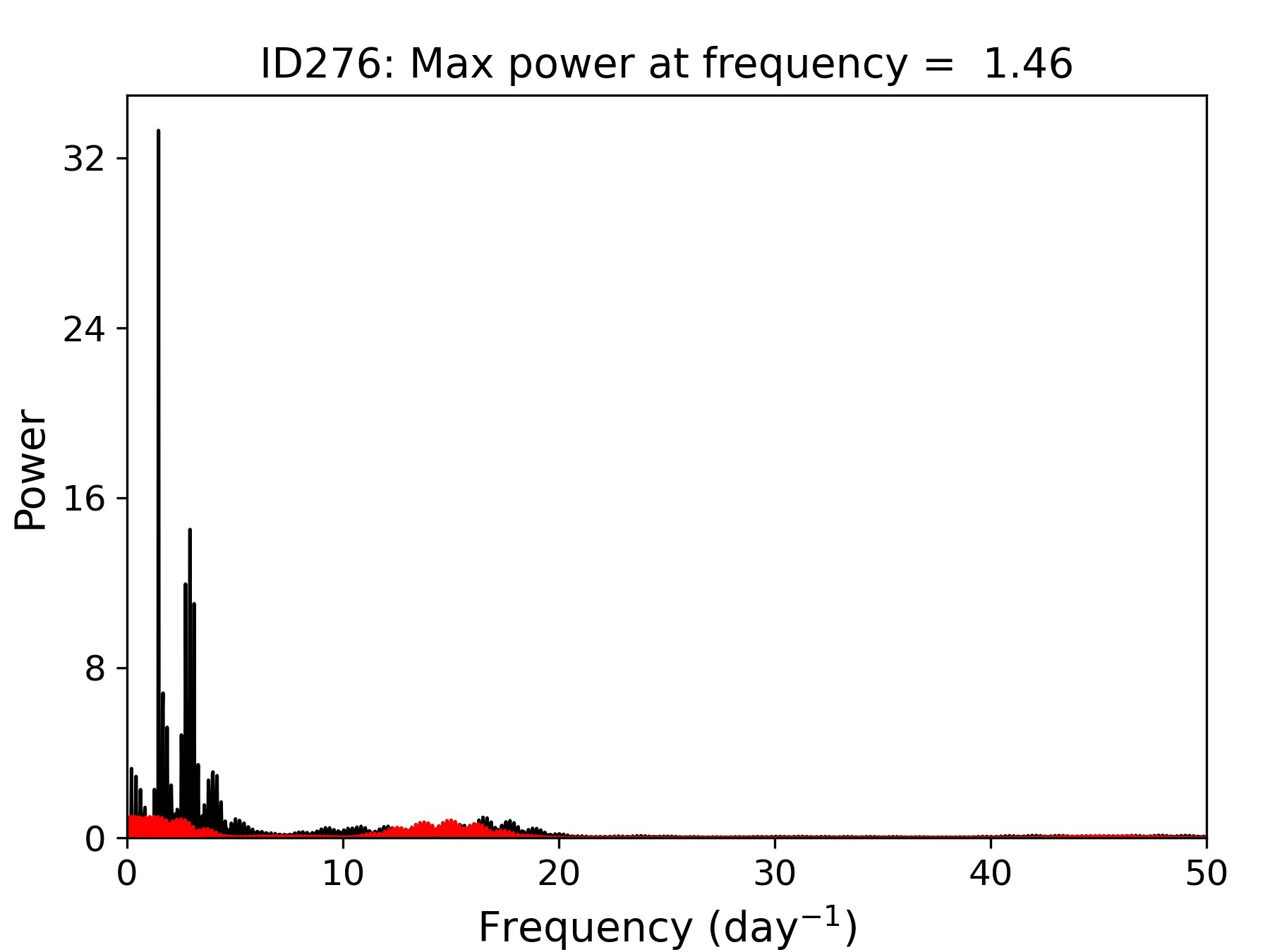}

    \caption{Lomb--Scargle periodograms (black) of the sources with periods measured along with the window function plotted in red.}
    \label{periodograms}
\end{figure}

\label{lastpage}

\pagebreak

\end{document}